\documentclass[graybox]{svmult}

\usepackage{type1cm}        
\usepackage{makeidx}         
\usepackage{graphicx}        
\usepackage{multicol}        
\usepackage[bottom]{footmisc}

\usepackage{newtxtext}       %
\usepackage[varvw]{newtxmath}       

\makeindex             

\usepackage{natbib}    

\def\kms{\rm\, km\,s^{-1}}
\def\msun{\,{\rm M_\odot}}
\def\ergs{\rm\, erg\,s^{-1}}
\def\mbh{\ensuremath{M_{\rm BH}}}
\def\fedd{\ensuremath{f_{\rm Edd}}}
\def\mstar{\ensuremath{M_{\rm star}}}

\begin{document}

\title*{Theoretical Modelling of Early Massive Black Holes}
\author{Marta Volonteri}
\institute{Marta Volonteri \at Institut d’Astrophysique de Paris, UMR 7095, CNRS and Sorbonne Universit\'e, 98 bis Boulevard Arago, 75014 Paris, France, \email{martav@iap.fr}}
%
%
\maketitle

\abstract*{This Chapter reviews theoretical models of massive black hole formation, growth and observables. It starts with a brief summary of basic properties of massive black hole properties. It then summarizes the current view on massive black holes and active galactic nuclei at high redshift, highlighting the JWST ``revolution'' and the questions raised by the recent observations. The Chapter then touches on massive black hole formation and growth mechanisms, emphasizing the processes at play at early cosmic times. It then reviews techniques for modeling the cosmic massive black hole evolution, with an emphasis on cosmological simulations, before approaching how observables are derived from models. It concludes with a section reflecting on the main questions on the JWST-discovered population in light of the material presented in the earlier sections.}

\abstract{This Chapter reviews theoretical models of massive black hole formation, growth and observables. It starts with a brief summary of basic properties of massive black hole properties. It then summarizes the current view on massive black holes and active galactic nuclei at high redshift, highlighting the JWST ``revolution'' and the questions raised by the recent observations. The Chapter then touches on massive black hole formation and growth mechanisms, emphasizing the processes at play at early cosmic times. It then reviews techniques for modeling the cosmic massive black hole evolution, with an emphasis on cosmological simulations, before approaching how observables are derived from models. It concludes with a section reflecting on the main questions on the JWST-discovered population in light of the material presented in the earlier sections.}

\section{Introduction}
\label{sec:intro}

These notes follow as faithfully as possible what was presented at the  54th Saas-Fee Advanced Course ``Galaxies and Black Holes in the First Billion Years as seen with the JWST''. Neither the lectures nor these notes have the ambition of being complete or comprehensive: my goal has been to distill and develop the physical intuition for the processes involved in modeling early MBH evolution. This in turn implies that I had to make simplifications and approximations. Furthermore, while I have tried to be objective, they represent my personal view, albeit informed by reading hundreds of papers over the last 20 years. In the various Sections I highlight reviews that can help the reader, as well as many of the papers that have informed my understanding and developed the field. 

I also want to stress two additional important points. The first is that when I discuss theoretical modeling the words have to be interpreted as ``this is what models find'', rather than ``this is how the Universe behaves''. The second is that observational results -- and theoretical approaches -- are in flux: JWST has revolutionized our understanding of the first billion years of the Universe, but many results are still ``green'' and I expect they will develop over the years. I will be curious to see in ten years time how much of the theoretical and observational understanding has evolved.

\section{Massive black hole basics}
\label{sec:basics}

Astrophysical black holes (BHs) have been observed mainly in two mass regimes: \emph{stellar black holes}, which are the remnants of massive stars and have masses up to a few tens, or perhaps hundreds, of solar masses, and \emph{massive black holes} (MBHs), whose origin is less clear and have masses from about $10^4 \msun$ to more than $10^{10} \msun$. We here focus on the latter, exploring their origin, growth and relation to the former. We also consider that BHs exist in the intermediate mass range: although observational evidence is still limited, but not null \citep{mezcuaIntermediatemass2017,greeneIntermediateMassBlackHoles}, their existence is generically predicted theoretically, linking low- and high- mass BHs.

MBHs are found in the centers of all massive galaxies in the local Universe, and in a fraction of dwarf galaxies \citep{greeneIntermediateMassBlackHoles}. Accretion of mass onto MBHs is the source of power in quasars and Active Galactic Nuclei (AGN). Quasars are the most luminous among AGN, with luminosities reaching $10^{47} \ergs$, more than $10^{13} \,{\rm L_\odot}$: a luminosity comparable to that of an entire galaxy, but produced in a region comparable in size to the solar system. In contrast to accreting, active MBHs, we also observe quiescent MBHs, with very little accretion taking place. Many MBHs in the local Universe are quiescent, and the most quiescent MBH for which we could measure an accretion rate is Sgr A$^*$, the MBH at the center of the Milky Way. The level of accretion is often described by the parameter \fedd, which compares the accretion luminosity to the Eddington luminosity: 
\begin{equation}
    \fedd=\frac{L}{L_{\rm Edd}}=\frac{\epsilon \dot{M} c^2}{4 \pi G \mbh m_p c/\sigma_T},
\end{equation}
where $\dot{M}$ is the accretion rate onto the BH, $G$ is the constant of gravity, $\mbh$ is the BH mass, $m_p$ is the proton mass,  $c$ is the speed of light and $\sigma_T$ is the Thomson cross section. $\epsilon$ is the radiative efficiency that quantifies the amount of rest-mass energy that goes into luminosity, and is generally related to the MBH spin (with modifications for highly sub-Eddington or super-Eddington accretion). 

In local galaxies MBH masses correlate with properties of the host galaxies, such as the bulge mass \citep{magorrianDemographyMassiveDark1998}, the stellar velocity dispersion \citep{ferrareseFundamentalRelationSupermassive2000,gebhardtRelationshipNuclearBlack2000}. A weaker correlation is also seen with the total stellar mass \citep[\mstar,][]{reinesRelationsCentralBlack2015,greeneIntermediateMassBlackHoles}. At high masses, where bulge-dominated galaxies dominate, bulge mass is a good predictor of MBH mass \citep{kormendyCoevolutionNotSupermassive2013}; the stellar velocity dispersion is the best overall predictor across the mass range \citep{nguyenImprovedDynamicalConstraints2019}, while the relation between MBH and total \mstar\ has a larger scatter but has the advantage \citep{reinesRelationsCentralBlack2015} that it is the ``easiest'' to measure – although at high redshift even \mstar\ is hard to pin down! See Richard Ellis's  Lecture 2, Section 3.3 that directly discusses this.

\begin{figure}
\includegraphics[width=\textwidth]{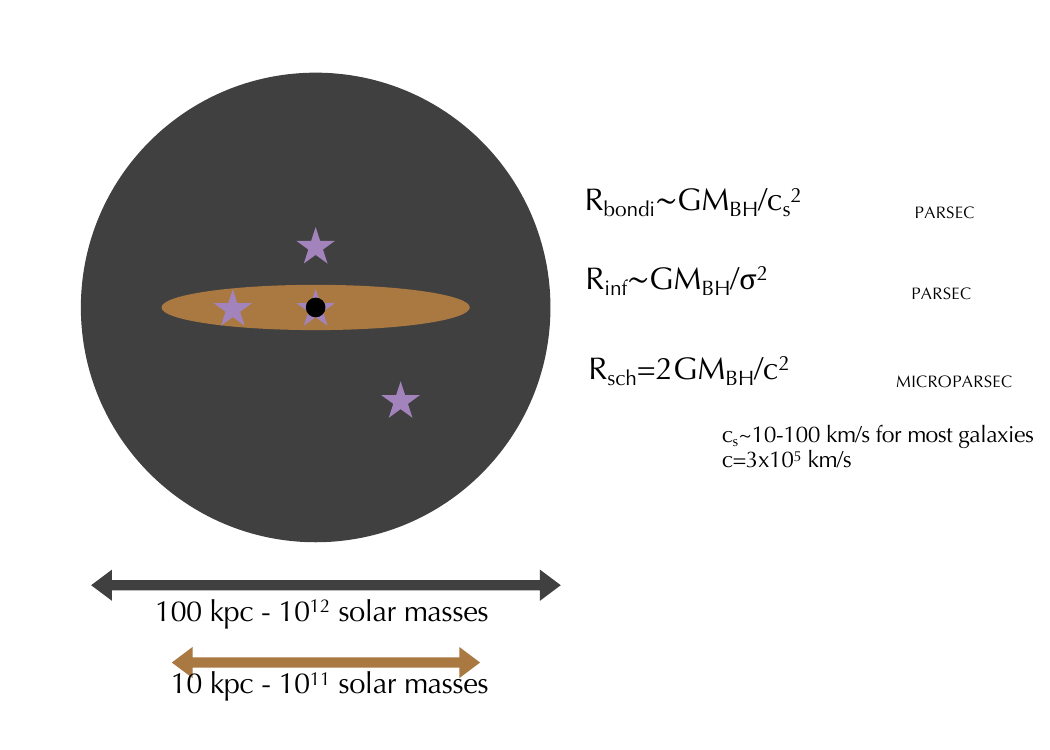}
\caption{Sketch of the physical scales characterizing galaxies and MBHs.}
\label{fig:scales}       
\end{figure}

It is instructive to compare the physical scales of MBHs in relation to galaxies (Fig.~\ref{fig:scales}). For a Milky Way-like galaxy, the dark matter halo has a mass of about $10^{12} \msun$ and a size of about 100~kpc. The stellar+gas component has a mass of about $10^{11} \msun$ and a size of about 10~kpc. For MBHs, we can consider three physical scales. The first is the Bondi radius, which corresponds to the region where gas is bound to a BH: 
\begin{equation}
R_{\rm Bondi}=\frac{2 G \mbh}{c_s^2},
\end{equation}
where $c_s$ is the sound speed. The sound speed is typically a few tens to hundreds of $\kms$, therefore the Bondi radius is of on scales of $\sim$~parsec. The second is the radius of the sphere of influence, marking the region where stellar dynamics is governed by the potential of the MBH. It is often\footnote{Another definition is the radius containing twice the MBH mass in stars. The definitions are equivalent for an isothermal sphere.} defined as  
\begin{equation}
R_{\rm inf}=\frac{2 G \mbh}{\sigma^2},
\end{equation}
where $\sigma$ is the stellar velocity dispersion, which also has values of a few tens to hundreds of $\kms$ for most galaxies, making $R_{\rm inf}$ also of the order of a few to tens of parsecs. Finally, we can define the gravitational radius:
\begin{equation}
R_{\rm g}=\frac{G \mbh}{c^2}.
\end{equation}
$R_{\rm g}$ is half of the Schwarzschild radius, and is of the order of $10^{-6}$~pc. The difference in scale between galaxies and MBHs makes theoretical modeling challenging: to model even only one MBH+galaxy system, one would have to cover 8 orders of magnitudes in spatial and time scales!

\section{JWST's high-redshift AGN and Little Red Dots}
\label{sec:JWST}

Theoretical works have been predicting the existence of numerous AGN at high redshifts ($z>6$) for many years \citep[see][for a compendium of various models]{2019MNRAS.485.2694A,2022MNRAS.509.3015H}. The theoretical AGN luminosity functions have generally been overproducing AGN with respect to extrapolations of the observational luminosity functions, and we generally tried to \emph{reduce} MBH growth in models to reconcile them with observations \citep[see discussions in, e.g.,][]{2015MNRAS.452..575S,2017MNRAS.468.3935H}. 
Most of the observational work, however, remained focused on bright quasars at these redshifts, and the lore on the observational side was that ``there are no AGN in high-z galaxies''. For many people therefore it came as a surprise that JWST easily detected AGN at high-z ($z=4-8$): broad-line emitters have been found in large numbers in many JWST surveys \citep{2023ApJ...954L...4K,harikaneJWSTNirspecFirst2023,maiolinoJADESDiversePopulation2023,mattheeLittleRedDots2024}. Many of these sources are  characterized by red colors and compact sizes, and were therefore dubbed Little Red Dots (LRDs). The Spectral Energy Distribution (SED) of LRDs is characterized by a V-shaped  with both a blue and a red excess, leading to ad-hoc color selection \citep{labbeUNCOVERCandidateRed2023,greeneUNCOVERSpectroscopyConfirms2024}. Many broad line emitters have however standard AGN SEDs \citep{maiolinoJADESDiversePopulation2023} and a further population has been identified via narrow emission lines \citep[indicating obscured AGN, e.g.,][]{scholtzJADESLargePopulation2023}. See Richard Ellis's Lecture 2, Section 5.1.

\begin{figure}[b]
\includegraphics[width=0.5\textwidth]{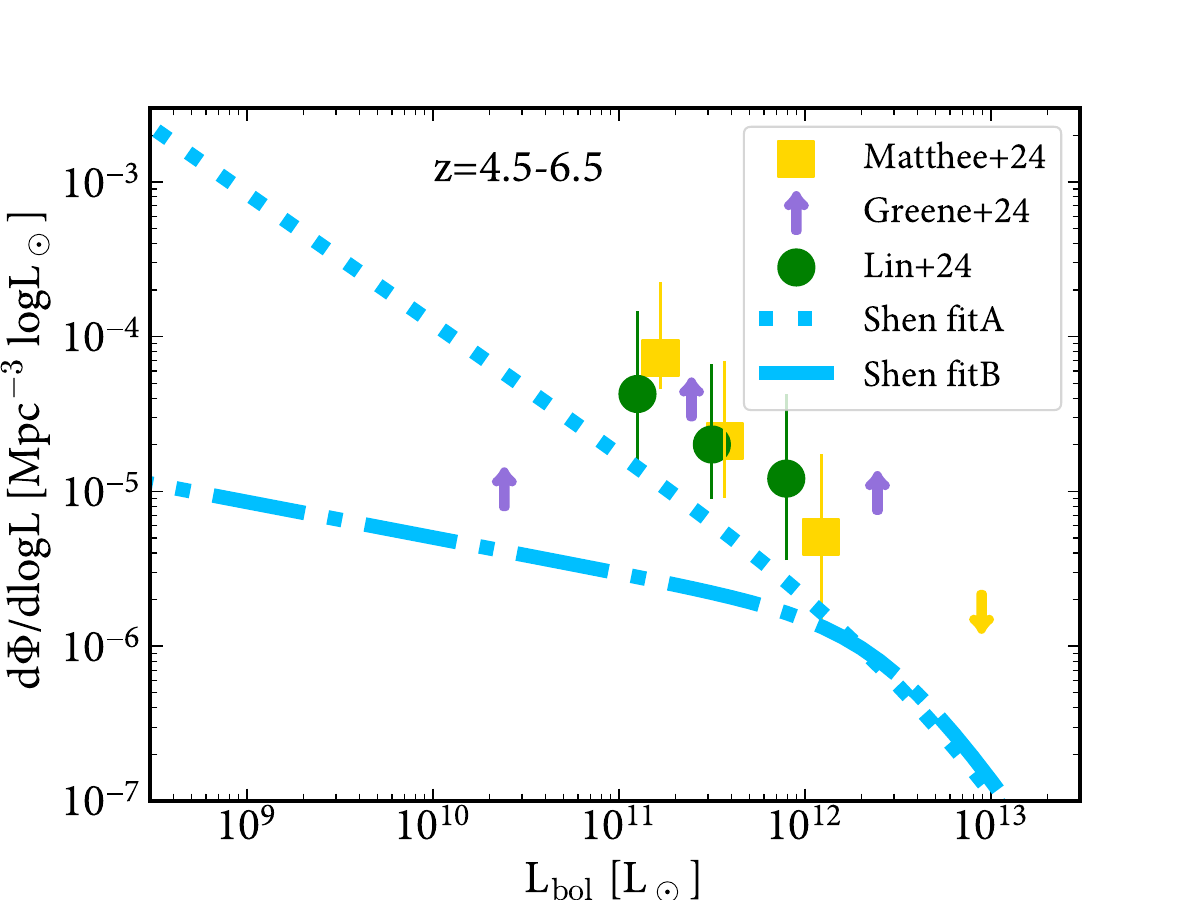}
\includegraphics[width=0.5\textwidth]{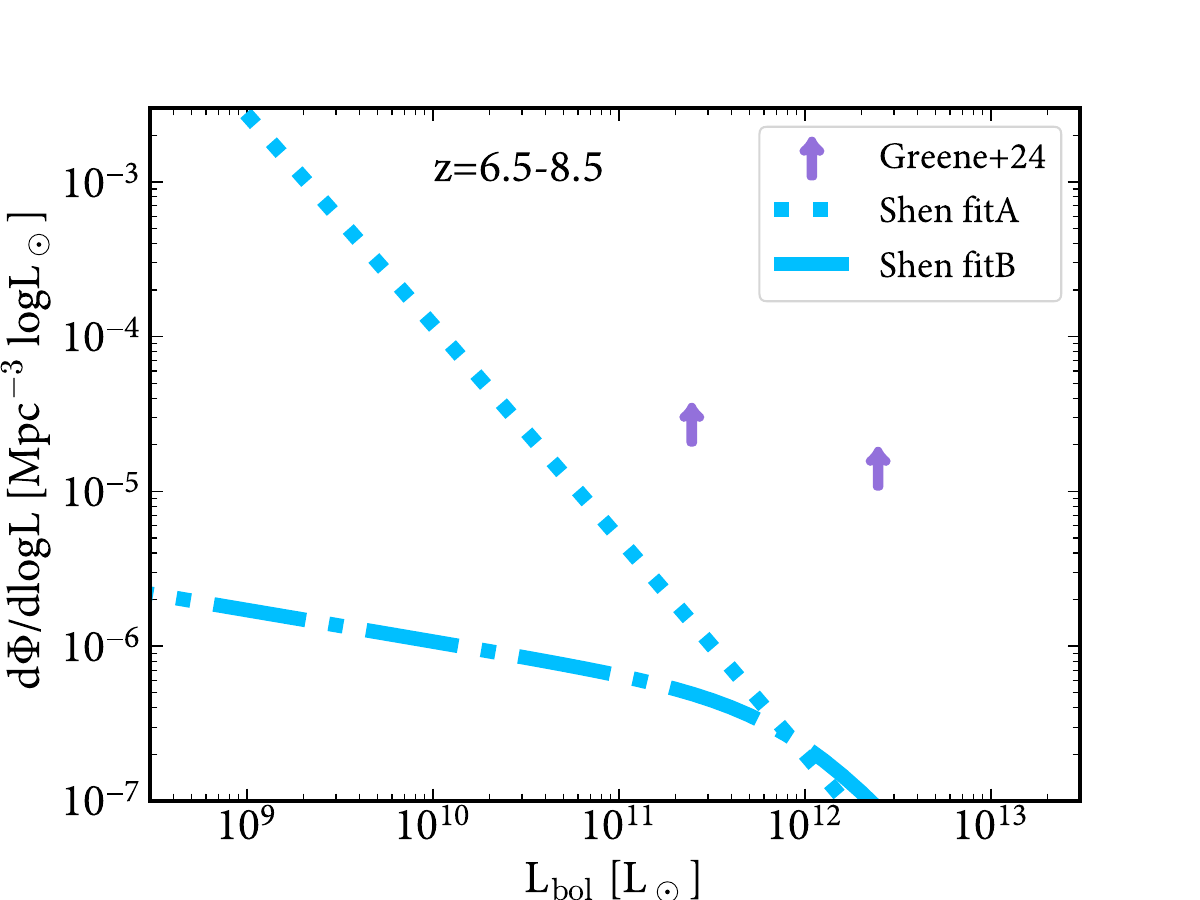}
\caption{AGN luminosity functions derived from JWST data \citep[yellow, green and purple points,][respectively]{mattheeLittleRedDots2024,2024ApJ...974..147L,greeneUNCOVERSpectroscopyConfirms2024}, compared to pre-JWST expectations \citep[blue curves,][]{2020MNRAS.495.3252S}. Standard relations based on local sources have here been used to estimate the bolometric luminosities of high-z AGN.}
\label{fig:AGNLF}       
\end{figure}

The fraction of galaxies hosting at $z=4-8$ AGN has been reported to be between 1\% and 20\%, in dependence of the selection criteria and the parent sample. I find particularly interesting the searches for broad line emitters in the EIGER, FRESCO and ASPIRE surveys, where the NIRCam/WFSS mode offers medium-resolution ($R\sim1600$) spectroscopy for all sources in the field, making a demographic analysis simpler. In contrast, JADES offers low- and high-resolution ($R\sim100$ to $\sim2500$) spectroscopy with higher sensitivity and larger wavelength coverage, but only for selected sources (due to the NIRSpec/MSA mask design), while UNCOVER targets sources with NIRSpec/MSA magnified by gravitational lensing: in both cases estimating completeness is more complex. In Fig.~\ref{fig:AGNLF} the bolometric luminosity functions are shown, in comparison with pre-JWST expectations, highlighting the enhanced number density of AGN with respect to extrapolation of pre-JWST data. As a note of caveat, these estimates assume that standard bolometric corrections apply to the high-z AGN, most of which are LRDs, which have unusual SEDs and may require different approaches \citep{2025arXiv250905434G}. Something to keep in mind is that for bright UV-selected quasars the searches completeness are well characterized: we are now seeing a completely different(-looking) population of AGN with JWST (they would not be color-selected by quasar searches, even if they were brighter). When we consider the bolometric luminosity function, instead, then obscured quasars and type 2 AGN could actually change the bright-end where quasars reside.

For high-z AGN with broad emission lines, standard techniques for MBH mass (see Eduardo Banados's Chapter) and galaxy (see Richard Ellis's Chapter) measurements have been generally applied to infer the relation between MBH and galaxy mass. The ratio of MBH to galaxy mass is elevated with respect to local sources \citep{maiolinoJADESDiversePopulation2023}: this was expected by some of us, because of selection biases \citep{Lauer2007b}. In simple terms, given the shape of the galaxy mass function, when selecting by AGN luminosity (e.g., the luminosity should be sufficiently high to power broad lines, or to make the AGN outshine the galaxy in a given band, or to skew the colors) it is more likely to select overmassive MBHs in common low-mass galaxies than undermassive MBHs in rarer massive galaxies. This led me to predict in a series of papers that high-z AGN would appear overmassive and therefore care should be taken when inferring scaling relations and demographics \citep{volonteriAssessingRedshiftEvolution2011,volonteriHighredshiftGalaxiesBlack2017,volonteriWhatIfYoung2023}. Whether the underlying relation is elevated \citep{2023ApJ...957L...3P} or not \citep{liTipIcebergOvermassive2024} remains highly debated\footnote{I highly recommend avoiding to combine a scatter plot of a sample (e.g., at high-z) with only the mean and scatter of another sample (e.g., the $z=0$ sample), but either all samples as scatter plots or as mean and scatter, and to always perform a statistical analysis to assess the importance of selection biases.}. An elevated relation, meaning that the MBH population is overall overmassive with respect to the local relation, implies a larger number of bright quasars compared to a non-evolving relation, therefore if there is confidence that the bright end of the quasar luminosity function has not been underestimated, it would imply that MBHs cannot be too overmassive. Conversely, discovering many quasars that have been missed by standard selection techniques would point to an actual overmassive population. 

\begin{figure}
\includegraphics[width=\textwidth]{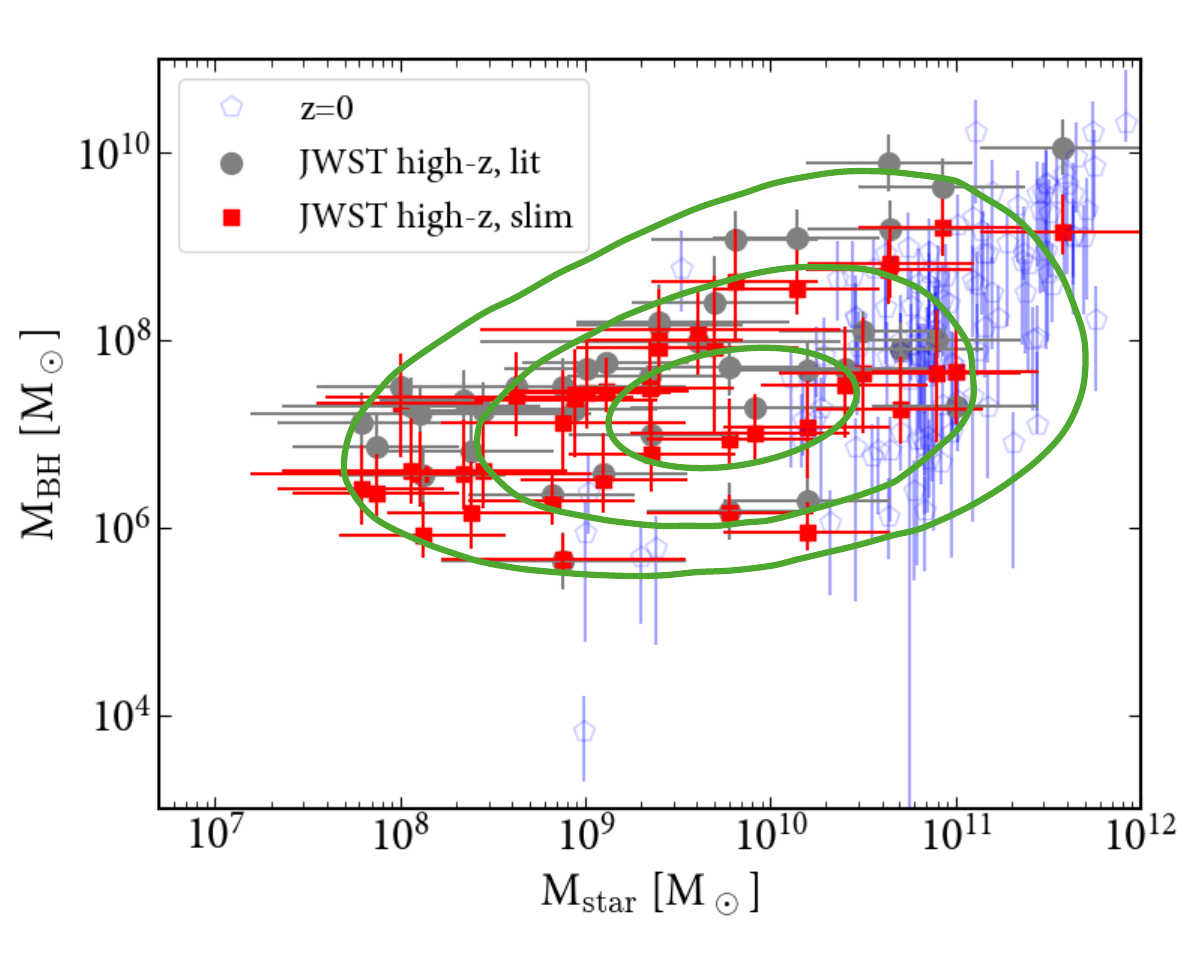}
\caption{Relation between \mbh\ and host galaxy \mstar\ for $z=0$ sources \citep[blue pentagons,][]{greeneIntermediateMassBlackHoles}, high-z AGN using \mbh\ from the discovery papers (gray dots) and including modifications to account for broad line region size in super-Eddington sources \citep[red, all data from][assuming a minimum uncertainty in \mstar\ of 0.45 dex]{2024A&A...689A.128L}. The green contours show the region where a MBH population which intrinsically sits on the $z=0$ relation from \citet{greeneIntermediateMassBlackHoles} is shifted when applying selection effects \citep[adapted from][considering an uncertainty in \mstar\ of 0.45 dex]{liTipIcebergOvermassive2024}.}
\label{fig:MbhMstar}       
\end{figure}

An additional point of note is on the definition and measurement (or estimate) of galaxy and MBH masses. Many local correlations focus on bulge, rather than total, stellar mass, since as I've already noted, the correlation is clearer for bulges. Since in the case of high-z galaxies bulge-disc decomposition is often unfeasible and we don't know if the hosts are bulge-dominated, it would be best to compare to relations calculated using total stellar mass, and not selecting preferentially bulge-dominated galaxies \citep[e.g.,][]{greeneIntermediateMassBlackHoles}. Further, when ALMA was the only instrument allowing for an estimate of the host mass, a dynamical mass, obtained via the virial theorem measuring a line width and a physical size, was often used \citep{Wang2010,2018ApJ...854...97D}. Consistency in how the host galaxy is defined is however important! For instance \cite{Bentz18} show how the results can vary even just in dependence of the adopted mass-to-light ratio, and there is no local determination of the relation between MBH and dynamical mass. Finally, galaxy stellar mass estimates are uncertain and systematic uncertainties are often difficult to assess, and this is even more true for MBH masses. MBH mass measurements are reviewed in Eduardo Bañados Chapter, and I will note here only that faint AGN masses are often estimated using an empirical calibration connecting the luminosity of the H$\alpha$ line to the continuum luminosity at 5100 $\AA$ \citep{2005ApJ...630..122G}. The latter is used to estimate the size of the broad line region, assumed to be in virial equilibrium in the MBH potential, through reverberation mapping campaigns. Given the significant difference in equivalent width of the lines between local and high-z AGN, there are caveats to this \citep{2025arXiv250905434G}. Since the equivalent width is normalized to the continuum, and luminosity at 5100 is assumed to probe the continuum, then a larger equivalent width means that H$\alpha$ is stronger with respect to the continuum than if it were excited in the same way as in the low-redshift sources used to establish the correlation. Furthermore, possible re-calibration of the relations may be needed to account for scalings of the broad line region size with luminosity or \fedd\ \citep{2015ApJ...806...22D,2024A&A...684A.167G,2024A&A...689A.128L}. Fig.~\ref{fig:MbhMstar} summarizes various of the points discussed here.

\begin{figure}
\includegraphics[width=\textwidth]{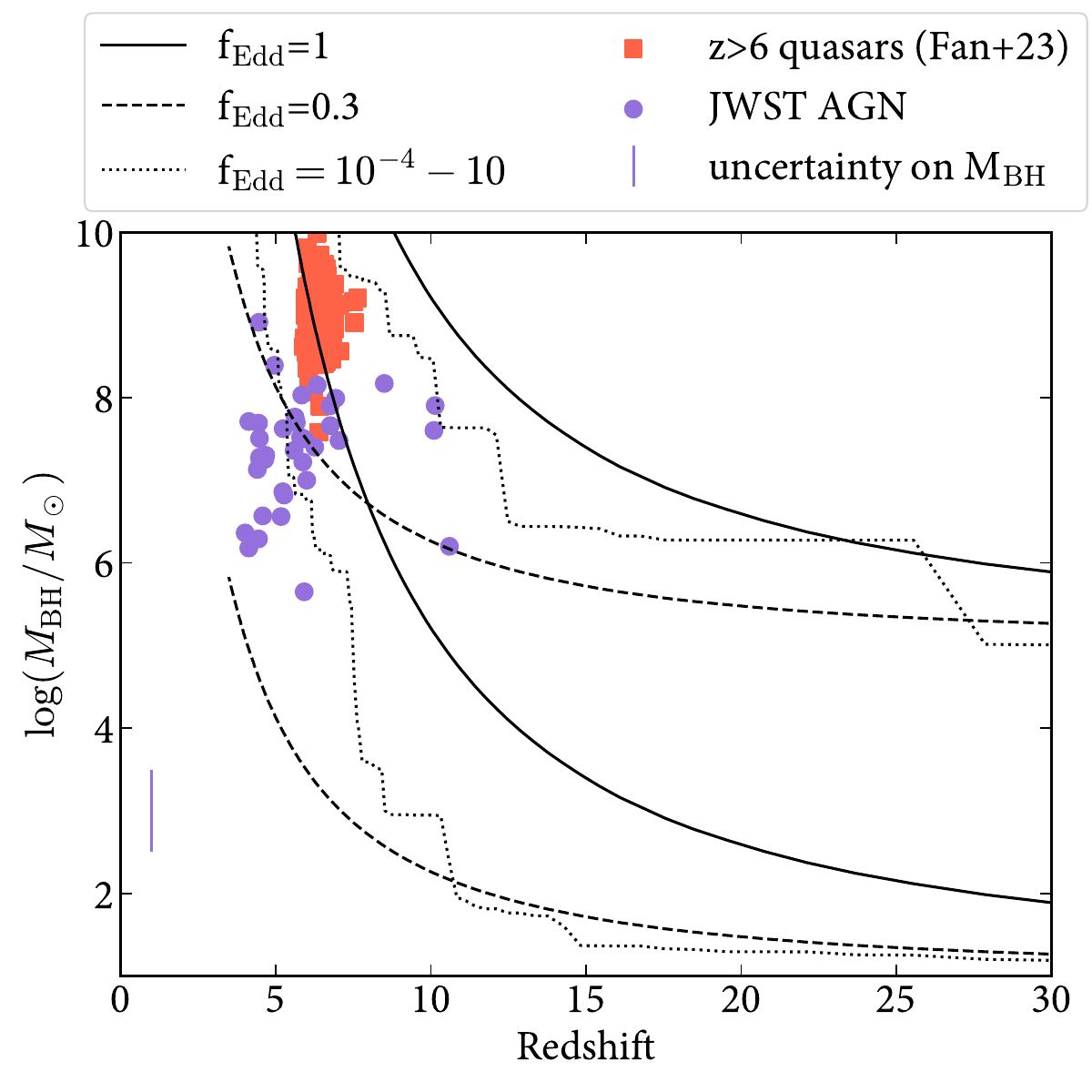}
\caption{Estimated MBH masses vs redshift for $z>6$ quasars \citep[red squares,][]{2023ARA&A..61..373F} and JWST-discovered MBHs at $z=4-11$ \citep[purple dots,][]{harikaneJWSTNirspecFirst2023,maiolinoJADESDiversePopulation2023,greeneUNCOVERSpectroscopyConfirms2024,2024NatAs...8..126B,2024ApJ...965L..21K}, compared to growth tracks assuming different initial MBH masses and accretion rates.}
\label{fig:MBH_redshift}       
\end{figure}

If the MBH masses are correct, they imply widespread and efficient MBH growth to reach such masses at such early times \citep[the same ``timing'' problem as for bright $z\sim 6-7$ quasars (see Eduardo Bañados's Chapter), but shifted to earlier times and lower masses, e.g.,][]{2023ApJ...953L..29L,Maiolino_2023}. This is exemplified in Fig.~\ref{fig:MBH_redshift}, where $z>6$ bright quasars (red squares) and $z=4-11$ AGN (purple) are shown in the $\mbh-z$ plane along with example tracks obtained integrating the equation:
\begin{equation}
\mbh(t)=\mbh(t_0) \int_{t_0}^t dt' e^{\frac{[1-\epsilon(t')]}{\epsilon(t')}\fedd(t')\frac{t'}{0.45 \rm{Gyr}}}.
\end{equation}
Here I have assumed $t_0=55$~Myr, $\epsilon(t')=0.1$, $\mbh(t_0)$ equal to $10 \msun$ or $10^5 \msun$, and three cases for \fedd: \fedd=0.3; \fedd=1 or $\log(\fedd)$ uniformly extracted from [-4,1) in timesteps of variable length from 2 to 15~Myr, to emulate variable accretion and duty cycles, including super-Eddington accretion. The latter exemplifies that including super-Eddington accretion allows for significant mass growth at a lower mean Eddington ratio ($\langle \log(\fedd) \rangle \sim-1.5$ in this example).

I will end this Section with some additional points on the ``weirdness'' of high-z AGN. The first bizarre fact is that the vast majority are X-ray weak: their X-ray luminosity is orders of magnitude below those of local AGN with similar masses and line/bolometric luminosities: \citep{2024ApJ...974L..26Y,2024ApJ...969L..18A,2025MNRAS.538.1921M}. 
2-10~keV observer's frame corresponds to $>10$~keV rest frame for $z>4$ sources, therefore only gas column densities above $10^{25} \,{\rm cm}^{-2}$ can significantly dim AGN. \cite{2025MNRAS.538.1921M} indeed suggest that the MBHs powering the high-z AGN are enshrouded by extremely dense gas within the dust sublimation radius. \cite{2024ApJ...976...96P} and \cite{2024arXiv240913047L} suggest instead that the AGN are super-Eddington accretors, where a combination of modifications in the SED and absorption decrease X-ray emission \citep[see also][for models linking super-Eddington and lack of X-rays]{2018ApJ...859L..20D,2017MNRAS.464.1102B,2024ApJ...976L..24M}. Finally, for, at least, LRDs there is a lack of reprocessed IR emission, suggesting low dust content despite the red colors \citep{2024ApJ...975L...4C,2025arXiv250522600C,2025arXiv250302059S}. I will return to these points in the rest of this Chapter, and in particular in Section~\ref{sec:before_after}.  On the other hand, high-redshift quasar properties, are at all wavelengths extremely similar to low-z quasars: at the bolometric luminosity overlap between ``high-z AGN'' and quasars there are distinct populations, or at least populations with distinct spectral features.

\begin{overview}{Overview}
To summarize, from high-z quasars we had already learned that at least some MBHs must grow efficiently in a short cosmic time. If all JWST AGN candidates are real AGN, and the luminosities/masses are correct, \textit{widespread} MBH formation and very efficient growth are needed to explain the population.  If MBHs, as a population, are overmassive, rather than the detectable population being biased, MBHs must outpace galaxy growth even in low-mass galaxies. 
\end{overview}

\section{MBH formation}
\label{sec:form}

In a seminal paper in 1978 Martin Rees outlined the ``possible models of formation of a massive black hole in a galactic nucleus'' \citep{1978IAUS...77..237R}. Starting from a gas cloud, he considered various pathways, which through different physical processes (collapse, accretion, stellar dynamical processes following star formation, etc). Today, we still consider the same models, with the addition that instead of starting from a generic gas cloud, we embed the initial conditions and environment in the cosmic evolution of structures. Besides these models that are based on gaseous and stellar processes, MBHs can also form before gas and stars, as, for instance primordial BHs. Much of what follows is inspired by two recent (pre-JWST) reviews \citep{2021NatRP...3..732V,2024OJAp....7E..72R}, and a comprehensive review is presented in \citet{2020ARA&A..58...27I}.  

\begin{figure}
\includegraphics[width=\textwidth]{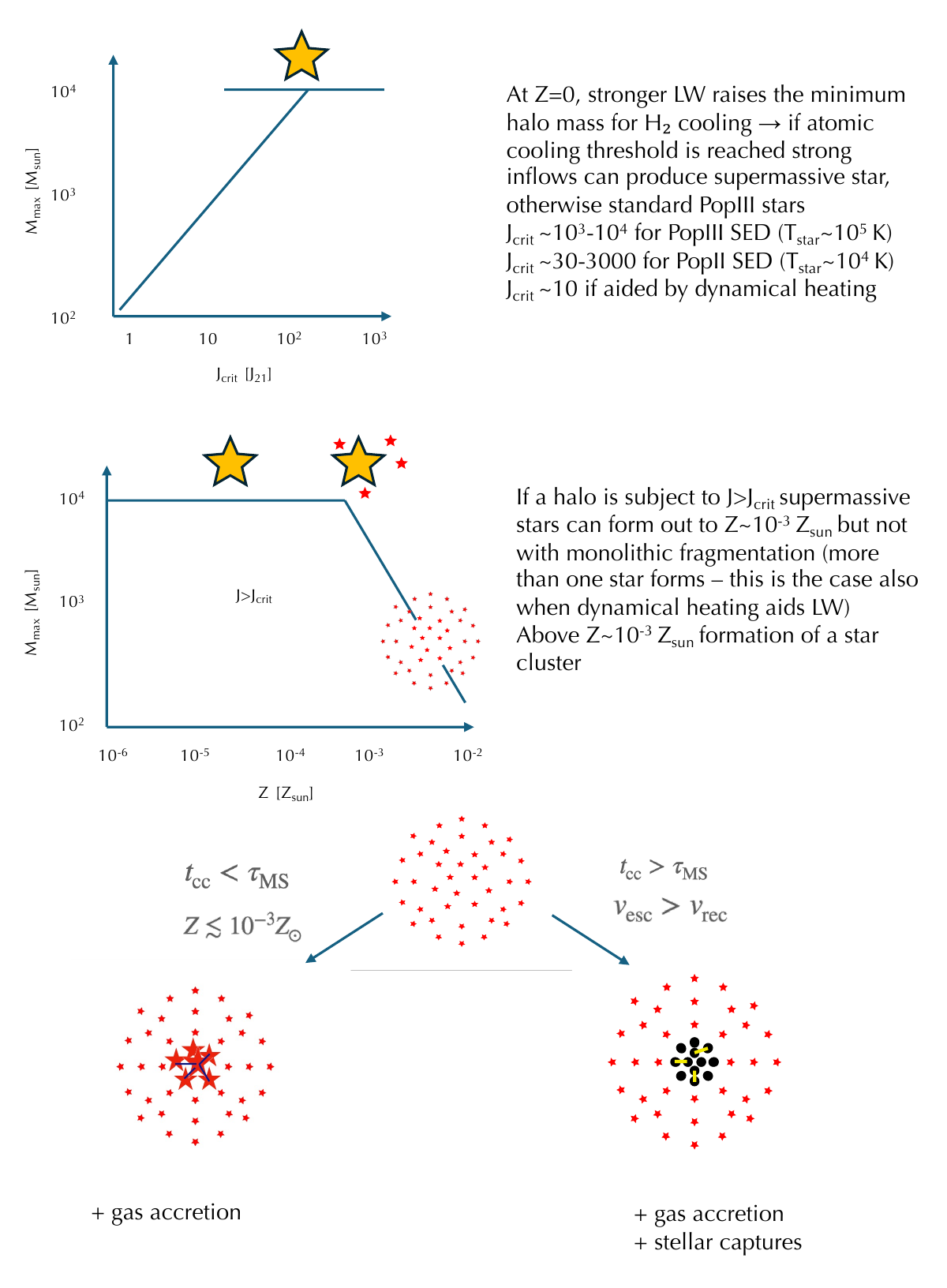}
\caption{Sketch of how different patways to MBH formation can be viewed as a continuum, in dependence of the environmental conditions: strength and SED of the LW radiation, metallicity, cluster properties.}
\label{fig:ContinuumSeeds}       
\end{figure}

Historically, MBH formation has started receiving attention in the early-mid 2000s. \citet{2001ApJ...551L..27M} suggested that MBHs would be the natural end-product of the first stars, and a few years later various papers explored formation of more massive `seeds', which were dubbed `direct collapse BHs' despite most of these models relying on an intermediate phase of a supermassive star or a quasistar \citep[][I will refrain from using ``direct collapse'' for these models here, as it can lead to confusion]{2003ApJ...596...34B,2006MNRAS.370..289B,2006MNRAS.371.1813L}. A third, less popular at the time, scenario considers runaway mergers of stars in dense stellar clusters \citep{2008ApJ...686..801O,Devecchi2009}. More recently, models suggest that \emph{instead of a bimodal behavior of light vs heavy seeds, there is more of a continuum, with the lightest and heaviest seeds as extremes of the distribution} \citep[see][for a discussion and relevant references, and Fig.~\ref{fig:ContinuumSeeds} for a schematic]{2024OJAp....7E..72R}. 
Many papers consider only the first two models and various ways to distinguish them have been proposed \citep{2008MNRAS.383.1079V,2018MNRAS.481.3278R,2018MNRAS.476..407V}. In some cases models have been combining a variety of formation mechanisms \citep{2010MNRAS.409.1022V,2014MNRAS.442.3616L,2021MNRAS.506..613S,2023MNRAS.518.4672S}, providing a more nuanced view of the relative importance of different scenarios.  

\subsection{Remnants of the first stars}
\label{sec:popIII}

The first stars form, by definition, when the only atomic elements are those produced by primordial nucleosynthesis. Their formation is therefore dominated by processes involving hydrogen in its various species. At zero metallicity the only coolants are H$_2$ and HD, which require photons and free electrons to form. 
H$_2$ can lower the temperature to $\sim 200$~K and can form even with only a small fraction of free electrons/photons. HD can lower the temperature to the CMB floor $2.7(1+z)$ but needs seed photons for forming deuterium, therefore it becomes effective only after ionizing sources exist. We generally distinguish between minihalos, which have a virial temperature $<10^4$~K, and atomic cooling halos, with higher virial temperature and masses of $\sim 10^7-10^8\, \msun$, where gas is partly ionized and formation of H$_2$ and HD is easier \citep{1997ApJ...474....1T}. 

There are two key points that make the first stars, also called PopIII, likely to end into relatively massive BHs ($\sim 100 \msun$). The first is that  the very first stars in minihalos have higher mass because the high temperature ($\sim 200$~K) implies a large Jeans mass:

\begin{equation}
    M_J= 500 \left(\frac{T_{\rm gas}}{200 \rm{K}} \right)^{3/2} \left(\frac{n_{\rm gas}}{10^4 \rm{cm}^{-3}} \right)^{1/2} \msun.
\end{equation}

The second key point about zero or low meallicity is related to stellar evolution.  No heavy elements in the composition of massive stars imply lower optical depth.
Pulsations/winds cause much less mass loss in the late-stage phases and as a consequence the mass of the remnant BH is a large fraction of the initial stellar mass \citep[e.g.,][]{2003ApJ...591..288H,2015MNRAS.451.4086S}. This is particularly relevant for stars above the pair instability gap\footnote{Stars in the pair instability gap, $\sim 50-140 \msun$, are expected to explode without leaving any remnant.}, $\gtrsim 140 \msun$, which collapse directly into a BH. 

However, in the early 2000s it was expected that each minihalo would form only one star, and that this star would be very massive, and stable in the center of its halo, thus the best conditions for collapse into a large BH and its further growth. That’s why it was considered a perfect way to seed MBHs \citep[e.g.,][]{2001ApJ...551L..27M,2003ApJ...582..559V}. 
Now, this is not considered the case anymore -- multiple stars are expected to form even in minihalos, with a mass
distribution that does not necessarily reach very high masses in each minihalo. This decreases the initial BH mass and thus hampers further growth. 

\subsection{Supermassive stars/quasi-stars}
\label{sec:SMS-QS}

The latter realization led some groups to consider more extreme stars: the formation of supermassive stars collapsing into MBHs of $\sim 10^5-10^6 \msun$. This is feasible if star formation is suppressed during the minihalo phase, and most of the gas in the still pristine atomic cooling halo is rapidly accreted
onto one central protostar. The condition can be expressed as a minimum critical accretion rate on a protostar: $\dot{M}>0.01-0.1 \msun \, \rm {yr}^{-1}$. This can be considered PopIII star formation at
its extreme. The requirements for this to happen is no star formation in the halo and primordial or very low metallicity \citep{2008ApJ...686..801O} gas composition, with further suppression of H$_2$ formation. Alternatively, turbulence or strong dynamical instabilities can prevent fragmentation and funnel gas in the halo center \citep{2006MNRAS.370..289B,2009ApJ...702L...5B,Mayer2015}. 

A first pathway to supermassive stars, up to $\sim 10^5-10^6 \msun$, but can be limited by mass loss \citep{2011MNRAS.417.3035D}, that has been explored much in the literature is via strong dissociating UV radiation in the Lyman-Werner band \citep[e.g.,][]{2003ApJ...596...34B,2013MNRAS.433.1607L,2014MNRAS.445.1056V,2014MNRAS.442.2036D,2016MNRAS.459.4209A}, thus preventing the formation of H$_2$ until the halo crosses the atomic cooling threshold \citep{Machacek2001}. The level of Lyman-Werner radiation needed to keep hydrogen fully dissociated until the halo grows enough and then a very massive star is formed is tens to thousands of times larger than the Lyman-Werner background, requiring nearby UV sources to enhance the local Lyman-Werner flux. The exact value of the critical intensity\footnote{In units of $J_{21}=10^{-21} \, \rm{erg^{-1} \, cm^{-2} \, Hz^{-1} \, sr^{-1}}$.} depends on the SED of the impinging radiation \citep{2011MNRAS.418..838W,2014MNRAS.445..544S,2015MNRAS.446.3163L}, and metallicity/dust content \citep{2006MNRAS.369.1437S}.  

In the classic metal-free case, this led to consider ``synchronized halos'' that remain pristine and cross almost at the same time the atomic cooling threshold, where the first one to form stars provides the UV radiation necessary to suppress H$_2$ in the other halo \citep{2014MNRAS.445.1056V,2016ApJ...832..134C}. Besides the low probability of synchronization, this scenario has two additional features making its success a rare process. The first is that after a few Myrs stars in the first halo start to explode and release metals, polluting the second halo. If the second halo becomes too\footnote{Strong UV radiation in metal poor, rather than metal-free, halos can lead to supermassive star formation in the presence of fragmentation if gas accretes preferentially on the most massive star \citep{2025MNRAS.539.2561C}.} metal enriched, then normal star formation takes place, instead of formation of a single supermassive stars. The window of opportunity for a supermassive star to form in the second halo is therefore small \citep{2016MNRAS.456.1901H,2016MNRAS.463..529H}. Pockets of metal free gas can persist if protected by dense filaments, but tidal stripping can dissolve the second halo as it falls into the main halo \citep{2016ApJ...832..134C,2021MNRAS.502..700C}. The process therefore appears to be rare \citep{2016MNRAS.463..529H,2024OJAp....7E..72R}. The requirement that the metallicity is exactly zero, however, may not be necessary, and supermassive star formation may occur up to $\sim 10^{-3} Z_\odot$ \citep{Chon-Omukai2020}, albeit with the formation of more than one star. Relaxing the requirement of zero metallicity can increase the number density, but the dynamical processes curtailing the evolution should still be at play. 

A second way to form very massive stars in pristine halos (likely less massive than in the case of strong UV radiation), is via ``dynamical heating'' (also known as rapid halo growth). In combination with moderate LW radiation, rapid growth of a halo can stave off star formation by injecting energy as the halo grows in mass and so does its virial temperature \citep{2003ApJ...592..645Y,2019Natur.566...85W}. In this case multiple stars form in the halo, reaching several thousands of solar masses \citep{2020OJAp....3E..15R}: rapid halo growth leads to less extreme BH masses. Stellar growth is terminated as stars consume available gas and move in and out of dense gas regions.

Finally, supermassive stars can be created by dynamical instabilities. There are at least four proposed mechanisms, where the first two also require some suppression of star formation. The first is via global instabilities, such as bars-within-bars \citep{2006MNRAS.370..289B}: a cascade of instabilities that transport angular momentum out and cause most of the gas to infall into a single object, perhaps a quasistar \citep[a “star” supported by accretion onto a black hole instead of nuclear fusion][]{2008MNRAS.387.1649B,2024ApJ...970..158C}. The second is via local (Toomre) instabilities \citep{2006MNRAS.371.1813L}: in a self-gravitating pre-galactic disc that is marginally stable gas infall can dominate over fragmentation. The third is major mergers between Milky Way-sized galaxies \citep{2010Natur.466.1082M}. Strong inflows can cause the formation of a $\sim 10^8-10^9 \msun$ clump that is so dense that it collapses into a BH under general-relativistic instabilities. Convergence of cold flows and turbulence can also help \citep{2022Natur.607...48L}. The first two mechanisms have somewhat lost favor because cosmological simulations don’t see these effects, while the last two appear to be too rare to explain the full MBH population \citep{2022Natur.607...48L,2023MNRAS.518.4672S}.

\subsection{Mergers of stars and/or BHs in star clusters}
\label{sec:dynchann}

As metallicity increases, star formation transitions to PopII, with an  extended initial mass function. At the same time, proto-galaxies become more massive, implying larger  masses of individual stars and multiple sites of star
formation, which is likely clustered: star clusters are born \citep{2008ApJ...686..801O,Devecchi2009}. Bound star clusters are in fact detected in high-z galaxies \citep{2019MNRAS.483.3618V,2023MNRAS.520.2180C,2024Natur.632..513A}, with some galaxies presenting steep central density profiles, reminiscent of nuclear star clusters \citep[e.g.,][]{2023ApJ...952...74T,2024RNAAS...8..207G,2024MNRAS.529.3301T}. 

In dense star clusters dynamical processes shape the population of stars and stellar remnants via core collapse, leading in some cases, at least theoretically,  to formation of a massive black hole \citep{Miller2002,2004Natur.428..724P,Gultekin2004,Freitag2007,2012ApJ...755...81M,Antonini2019}. 

Generically, the first step is mass segregation, i.e., the most massive stars sink towards the center of the cluster. As massive stars become confined in a smaller region, stellar interactions increase, and lead to energy exchange: stars that get kinetic energy leave the cluster, decreasing the cluster’s kinetic energy. To balance the loss of kinetic support, the cluster contracts and becomes denser. This in turn increases the rates of stellar interactions: in a runaway process the cluster becomes denser \citep[I stress that this are beautiful classical results, and I invite the reader to read about the developments in, e.g.,][]{2004cbhg.symp..138R}. Meanwhile, the most massive stars segregated in the center start to collide and merge, and as stars become more massive via collisions, their lifetime decreases. At this point the evolution bifurcates, depending on whether the core collapse time, $t_{\rm CC}$, is shorter or longer than the lifetime of the most massive star, $\tau_{\rm MS}$:

\begin{equation}
t_{\rm CC}\sim  3 \left( \frac{R_c}{\rm{pc}}\right)^{3/2} \left(\frac{M_c}{5\times 10^5 \msun} \right)^{1/2} \left( \frac{10 \msun}{\langle m \rangle} \right)\left(\frac{8.5}{\ln{\Lambda_C}} \right) \rm{Myr},
\end{equation}
where $R_c$ is the half-mass radius of the cluster, $M_c$ is the total mass of the cluster, $\langle m \rangle$ is the mean stellar mass, and the 
Coulomb logarithm $\ln{\Lambda_C}\sim \ln(0.1 M_c/\langle m \rangle)$ \citep{2002ApJ...576..899P}. 

If the runaway stellar merger phase causes the most massive star to grow above the pair instability gap before any other star explodes as supernova (SN), then a BH with mass $10^2-10^4 \msun$ can form, in dependence of the cluster mass $\mbh \sim 10^{-3} M_C \ln(\tau_{\rm MS}/t_{\rm CC})$ \citep{2002ApJ...576..899P}. Furthermore, metallicity had to be low $\lesssim 10^{-3}$ $\rm{Z}_\odot$, otherwise stars loose mass during collisions and via stellar winds \citep[e.g.,][]{2006MNRAS.366.1424D,2009A&A...497..255G,2016MNRAS.459.3432M}.  Indeed, collisions may foster mass loss instead of mass gain!

Let’s consider now the cases for which the core collapse time is longer than the lifetime of massive stars. This is the most common situation, as stellar binaries delay or even halt collapse\footnote{This is because in interactions between binaries and single stars the binaries get tighter and therefore provide kinetic energy to the cluster.} Massive stars explode as supernovae (SNae) and leave BHs and other compact objects (white dwarfs, neutron stars). Mass segregation remains at play, and therefore the most massive objects in the system, the BHs, sink to the center and ``decouple'' from the rest of the cluster. Within this BH subcluster, BHs couple in binaries via 3-body interactions or exchanges: interactions tighten BH binaries but at the same time eject some of them. When BH binaries become sufficiently tight, i.e., the BHs become very close, they merge via emission of gravitational waves (GWs). Emission of GWs causes a linear momentum imbalance, which causes the merged binary to recoil, with a speed that can reach $\sim 1000 \kms$ or more, in dependence of the binary mass ratio, spin and orbital configuration \citep{Lousto12}. If the cluster is sufficiently dense that the recoil velocity is less than the escape velocity from the cluster, the merged BH is retained. Since it is more massive than all other objects, it starts to grow hierarchically, merging with lighter BHs, as long as the recoil velocity is $<$ than the escape velocity. Typically, only a few mergers can occur before ejection. In contrast with the  runaway stellar merger case, the timescales for this process to build a massive BH are longer, but there is no metallicity dependence. 

On top of these two classical processes, stellar captures and gas
accretion can further increase the mass of the forming BH. If a cluster is left with only one BH, it cannot be ejected anymore, and it can grow by capturing stars up to perhaps $\sim 10^4-10^5 \msun$ \citep{Miller2002,2017MNRAS.467.4180S,2017NatAs...1E.147A,Rizzuto2021}. During the whole process, if gas is present, both stars and BHs can accrete gas and speed up the growth process  \citep{2014Sci...345.1330A,Boekholt2018,Chon-Omukai2020,Reinoso2018}. 

\subsection{Primordial BHs and cosmic string loops}
\label{sec:PBHs}

MBHs may form as early as at inflation: before galaxies \citep[e.g.,][]{2001JETP...92..921R,2015JCAP...06..007B,2018MNRAS.478.3756C}. In this case, the galaxy grows around the BH that provides the seed of the potential well for accreting dark matter and baryons. This eases MBH growth issues: MBHs stay put in the center at least for the early stages and can start accreting, although the feedback exerted by the BH may hamper its growth \citep{2008ApJ...680..829R}. 

Primordial BHs are dark matter candidates. They may form from the collapse of the very high tail of density fluctuations, or if there is a sudden change in the plasma pressure (e.g., phase transitions). The BH mass would be that of the horizon at the time of collapse, ranging from $\sim 1$~g at the end of inflation to $10^5 \msun$ at the time of pair annihilation \citep{Garcia-Bellido:2019tvz}. Several constraints limit the amount of PBHs at various mass scales, especially for monochromatic mass distributions \citep[see][for a recent review]{2020ARNPS..7050520C}. Early Universe constraints are based on Cosmic Microwave Background measurements, while late Universe constraints are based on the effect of BHs in galaxies or as sources of an X-ray background \citep{2022MNRAS.517.1086Z}. Primordial BHs with mass $<10^5 \msun$ at formation would be broadly consistent with early Universe constraints, with subsequent growth also consistent with late Universe constraints. The local number density of MBHs of $\sim 10^5 \msun$ is estimated to be about $10^{-3} \rm{Mpc}^{-3}$ \citep{greeneIntermediateMassBlackHoles}, and the total mass density at $z=0$ is about $5 \times 10^5 \msun \rm{Mpc}^{-3}$: they correspond to density parameters $\Omega_{MBH}\sim 10^{-10}$ and $10^{-7}$ respectively. 

Cosmic string loops are topological defects that are predicted by string theory and some quantum field theories. They can form during phase transitions in the early
Universe when cosmic strings intersect. If loops have sufficiently low angular momentum to contract within their own Schwarzschild radius, or if they are able to accrete enough dark matter, they can collapse into BHs. A priori masses and number densities can have a vast range, but Cosmic Microwave Background and gravitational waves place limits, which allow BHs with masses up to $10^5 \msun$ at $z\sim 30$ when requiring that most galaxies host MBHs sourced by loops \citep{2015JCAP...06..007B}. 

\begin{figure}
\includegraphics[width=\textwidth]{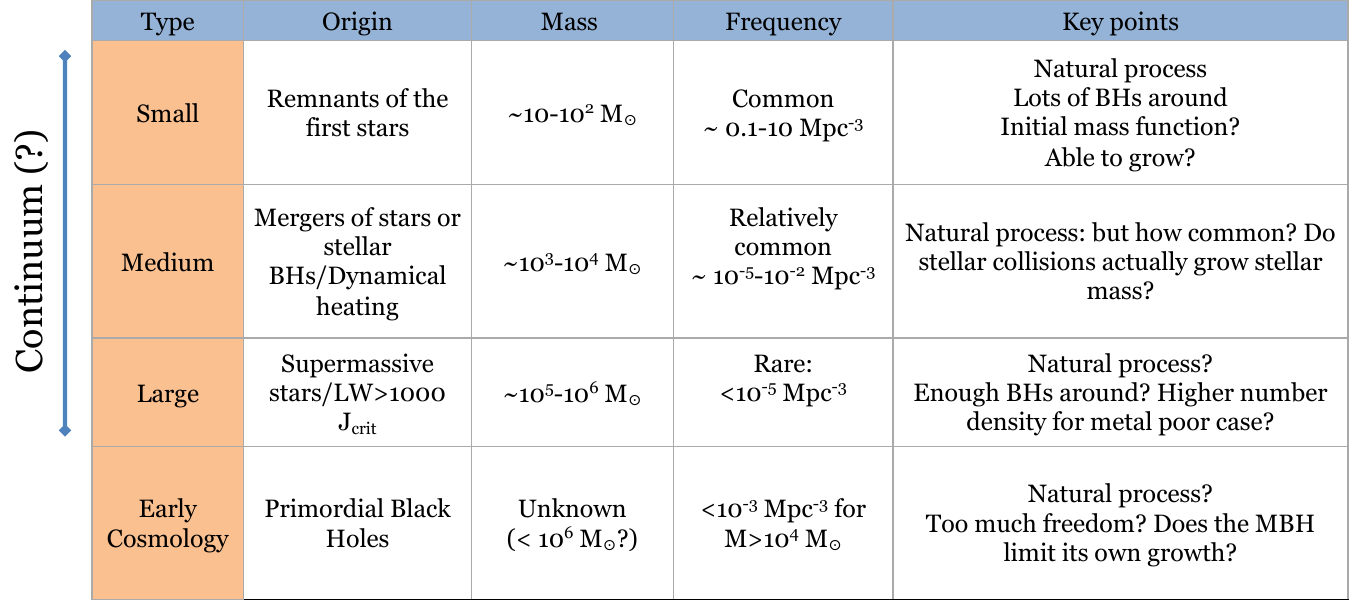}
\caption{Schematic view of MBH formation models and open questions}
\label{fig:SchemeSeeds}       
\end{figure}

\begin{overview}{Overview}
To summarize, several processes can contribute to forming MBH seeds. Processes related to gas and stars shouls be considered more of a continuum rather than a dichotomy between ``light'' and ``heavy'' seeds. The rare heaviest seeds are expected in strong UV radiation sites, more common situations lead to the formation of a multiplicity of lower-mass BHs, with relatively high mass at birth (PopIII stars) or gained by stellar or BH interactions in star clusters. See Fig.~\ref{fig:SchemeSeeds} for a summary of models and key questions. 
\end{overview}

\section{MBH Growth}
\label{sec:growth}

MBHs at birth are significantly less massive than the MBHs powering quasars and sitting in today's massive galaxies: both can exceed $10^{10} \msun$. MBHs grow in mass through three processes: accretion of gas, MBH-MBH mergers and accretion of stars. Each of these processes is related to observational probes: AGN, GWs and Tidal Disruption Events (TDEs). Not all growing MBHs can be detected, though: AGN must be sufficiently bright, meaning that the accretion rate cannot be arbitrarily low; GWs must fall in the frequency range of detectors and be sufficiently strong, TDEs don't accompany stellar accretion for the most massive MBHs, where the stellar tidal radius is within the Hill radius (of the order of the Schwarzschild radius for a non-spinning BH).

A beautiful argument developed by \cite{2002MNRAS.335..965Y} and building upon Soltan's argument \citep{1982MNRAS.200..115S}, allows one to estimate the relative importance of gas accretion with respect to mergers. Mergers just reshuffle the distribution of masses, but the total mass density in MBHs is constant in time\footnote{Note that when two MBHs merge, the final mass is a few percent smaller than the sum of the two masses, because part is lost in GWs \citep{2010CQGra..27k4006L}.}. Accretion instead adds external matter, so that the total mass density in MBHs grows with time. Soltan showed that by integrating the AGN luminosity function over luminosity one can obtain the total energy density emitted by accreting MBHs as a function of time:

\begin{equation}
    u_{\rm AGN}=\int_0^t dt \int_0 ^\infty dL \, \Phi(L,t) L,  
\end{equation}

where $\Phi \equiv dN/dL dV$ is the luminosity function. Since $L=\epsilon \dot{M} c^2$ and threfore only a fraction $(1-\epsilon)$ goes into growing the MBH mass, one can get  the total mass density accreted by MBHs:

\begin{equation}
    \rho_{\rm AGN}= \frac{(1-\langle \epsilon\rangle )}{\langle \epsilon\rangle c^2} \int_z^\infty \left| \frac{dt}{dz} \right|  \int_0 ^\infty dL \, \Phi(L,t) L,  
\end{equation}
with several caveats, in particular related to obscured accretion and the value of $\langle \epsilon\rangle$ (in principle one could also leave $\epsilon(L,t)$ inside the integral, if ways can be used to estimate the dependence on time and luminosity). If integrated over time, this can be compared to the total mass density in MBHs, via integration of the MBH mass function, showing relatively good agreement, implying that most of the mass in today's MBHs can be explained by accretion alone. One can perform the same exercise at different times, to obtain the evolution in the total mass density accreted by MBHs. \citet{2024ApJ...973L..49I} note that while the evolution from $z=3$ to $z=0$, derived using pre-JWST data, is smooth, the $z>3$  population of JWST-detected AGN has a completely different behavior: the total mass density accreted by MBHs at $z\sim 8$ is comparable to the total mass density accreted by MBHs by $z\sim 2$ if only the pre-JWST data alone is used! Various solutions have been proposed, from high MBH spins that increase $\epsilon$ \citep{2024ApJ...973L..49I} to adopted dust corrections to bolometric luminosities being too high \citep{2025arXiv250522600C}, to super-Eddington accretion lowering the duty cycle \citep{2024arXiv241214248T}. 

Do MBH mergers and stellar accretion have a role, then? While waiting for a full census of MBH mergers via GW observations \citep[current constraints are only for the most massive black holes at $z\lesssim 2$][]{2023ApJ...952L..37A,2024A&A...685A..94E,2023ApJ...951L...6R,2023RAA....23g5024X} and of stellar accretion via TDEs \citep{2016MNRAS.455..859S}, we can consider theoretical predictions. Models suggest that other processes dominate growth ``when there is no gas available'', for instance MBH mergers contribute a large fraction of the mass in MBHs in massive, gas poor galaxies \citep{2014MNRAS.440.1590D,2015ApJ...799..178K}, while stellar accretion can contribute most to MBHs in low-mass galaxies where accretion is inefficient \citep{2021MNRAS.500.3944P,2024A&A...689A.204P}. 

\subsection{Gas accretion and Feedback}
\label{sec:feedback}

Feedback was first introduced in relation to star formation. Stars form in gas clouds where gas is very dense and cold, but stars emit ultraviolet light and then explode as SNae: the energy ``given back'' heats and rarefies gas preventing
further star formation. The same occurs in the case of MBHs, which preferentially accrete dense and cold gas, but inject energy in various forms in their surroundings. The energy “given back” when an MBH accretes gas also heats and rarefies the gas preventing further MBH growth (and modulating star formation in the galaxy). 

We can consider a vicious -- or virtuous -- feeding/feedback cycle. The galaxy feeds the MBH through gas inflows, activating it. An active MBH launches winds or jets that interact with the gas of the galaxy, modulating gas accretion onto the MBH and the formation of stars in the galaxy.  If gas is thus prevented from flowing to the MBH, the active MBH goes back to quiescence.  Gas cools and can form stars and feed the MBH: the cycle restarts. 

Typically, it is considered that stellar feedback (e.g., SN explosions) is the dominant source of feedback in dwarf galaxies, and AGN feedback takes over in massive galaxies \citep{2012RAA....12..917S}. AGN feedback may, however, have a role also in dwarf galaxies \citep[][and references therein]{2017ApJ...839L..13S,2022MNRAS.516.2112K}. 

SN explosions may have also an impact on MBH growth. Many simulations find that in dwarf galaxies SN feedback is also able to suppress star
formation and MBH accretion \citep[e.g.,][]{2015MNRAS.452.1502D,2017MNRAS.465...32B,2017MNRAS.472L.109A}. In the shallow potential wells of dwarf galaxies, SNae can cause rapid, dramatic fluctuations in the gas density near the MBH \citep{2017ApJ...836..216P}, which limit gas accretion onto the MBH. This phenomenon is reduced in the presence of nuclear star clusters, which deepen the potential well \citep{2025MNRAS.537..956P}. While the role of SNae in suppressing MBH growth is debated, if MBHs grew significantly in dwarf galaxies, the faint end of the AGN luminosity function would be overestimated \citep{2017MNRAS.468.3935H,2022MNRAS.511.5756T}. The number densities of high-z AGN found by JWST challenge this picture, because high-mass active MBHs are found in low-mass hosts (but see the caveats discussed in Section~\ref{sec:JWST}). A key question is therefore if AGN have been missed also at lower redshift: if we are confident that the AGN demography at $z<4$ is correct, then we must understand what changes in MBH evolution in dwarf galaxies between $z=2-3$ and $z>4$.

\begin{figure}
\includegraphics[width=\textwidth]{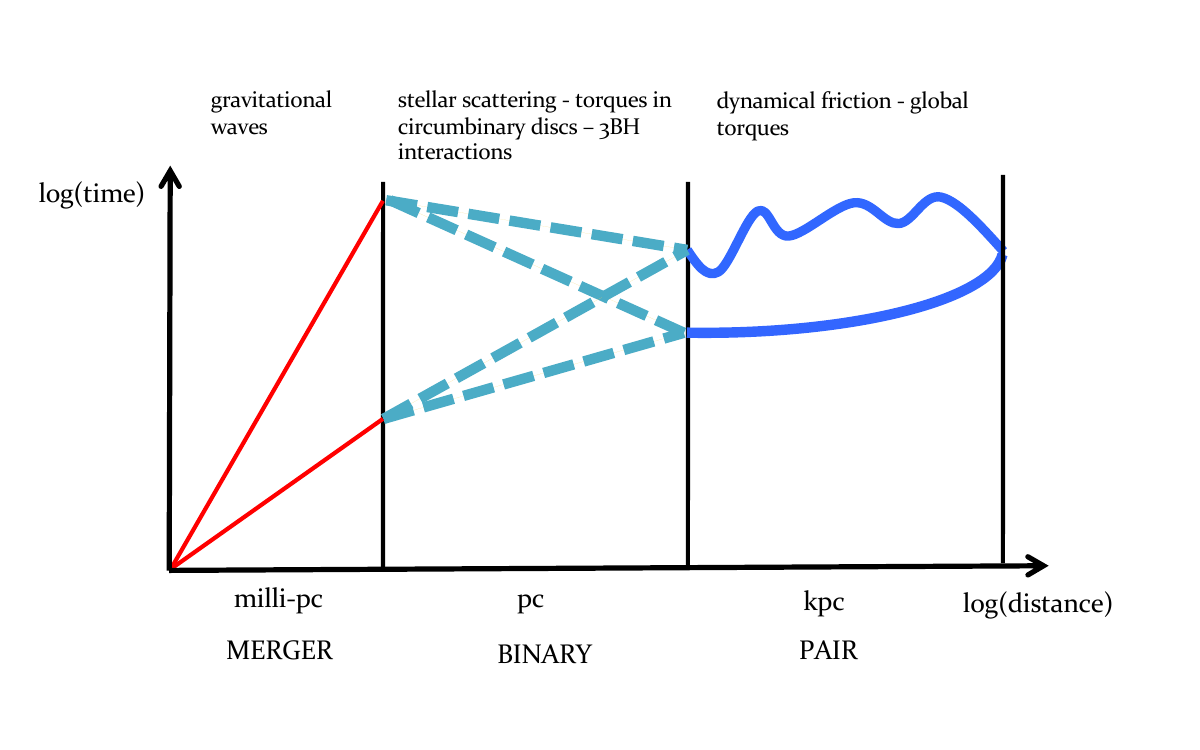}
\caption{Schematic view of scales and physical processes characterizing the dynamical evolution of MBHs following galaxy mergers.}
\label{fig:SchemeDynamics}       
\end{figure}

\subsection{MBH mergers}
\label{sec:mergers}

MBHs are expected to grow along with galaxies through accretion and MBH-MBH mergers. I refer the reader to Section~2.1 in \citet{2023LRR....26....2A} for a complete and exhaustive review of the (astro)physics of MBH dynamics and mergers. 

Studying MBH mergers is an extremely complex problem because of the scales involved. Besides the galaxy-scale problem already noted in Section~\ref{sec:basics}, we have to embed the problem in the cosmological context, because MBH mergers are sourced by mergers of galaxies containing (at least) one MBH each. This in turn requires the galaxy merger rate (and the occupation fraction of MBHs in galaxies). Models have therefore to probe scales of hundreds of Mpc and comprise a statistical sample of galaxies and MBHs. 

The journey of merging MBHs starts with the merger of the host galaxies, and is initially driven by its physics: the MBHs are simply a component of the merging galaxies, and usually their orbital evolution is driven by that of the stellar envelope (bulge or nuclear star cluster) in which they are embedded \citep{2002MNRAS.331..935Y}. In the case of mergers between galaxies with significantly different masses (mass ratios smaller than $\sim 0.1$) the disruption of the smaller galaxy can sometimes leave the MBH ``stranded'', although nuclear star formation in the secondary can reverse this situation for mass ratios of the order of 0.1 at least \citep{2009ApJ...696L..89C,2014MNRAS.439..474V}. 

The evolution is sketched in Fig.~\ref{fig:SchemeDynamics}, and was first described by \citet{1980Natur.287..307B}. Initially the MBHs evolve under the effect of dynamical friction, or global torques in the presence of bars. In a smooth potential dynamical friction causes orbital decay towards the
center, on timescales that for an isothermal sphere can be calculated analytically as \citep{1987gady.book.....B}:

\begin{equation}
 t_{\rm DF}=0.67 \left(\frac{a}{4 \rm{kpc}} \right)^2 \left(\frac{\sigma}{100 \kms} \right)   \left(\frac{\mbh}{10^8 \msun} \right)^{-1} \frac{1}{\ln \Lambda},
\end{equation}

where $a$ is the initial separation, and the term inside the Coulomb logarithm is often approximated as the ratio of galaxy to MBH mass, as a proxy for the maximum and minimum relevant impact parameters relevant for encounters between stars and the MBH. While galaxies are not all necessarily isothermal spheres, this equation gives a good feeling of the main dependencies: decay is longer for smaller MBHs in larger/massive galaxies. When dealing with high-z galaxies, their potential is often shallow and granular, making it difficult to define a center: the MBH dynamics can then become erratic, lengthening or hampering the formation of bound MBH binaries \citep{2019MNRAS.486..101P,Ma-Hopkins2021}. Global torques during the formation of bars have similar effects \citep{2020MNRAS.498.3601B}, and AGN feedback is also expected to play a similar role when dynamical friction is exerted by a gas distribution \citep{2017ApJ...838..103P,Souza-2017}. A natural corollary of long dynamical friction for light MBHs, difficult binding and erratic dynamics is the existence of wandering MBHs \citep{1994MNRAS.271..317G,2002ApJ...571...30S,2003ApJ...582..559V}, which are now found even at high-z \citep{2024MNRAS.531..355U}. 

When MBHs find each other and are able to bind into a binary, further evolution is driven by scattering with stars \citep{1996NewA....1...35Q}, migration in a circumbinary disc \citep{2002ApJ...567L...9A,2009ApJ...700.1952H} and triple interactions with additional MBHs if the residence time of the binary is sufficiently long \citep{2003ApJ...582..559V,2018MNRAS.477.3910B,2018MNRAS.473.3410R}. Eventually, when the MBHs get sufficiently close, emission of GWs, with a timescale scaling as the MBH separation to the fourth power, takes over and leads to coalescence \citep{1964PhRv..136.1224P}. MBHs embedded in nuclear star clusters may have a much easier life throughout this process, as the relevant mass for dynamical friction will be the cluster mass, and stellar hardening is more effective  \citep{2020MNRAS.493.3676O,2025ApJ...981..203M}.

\subsection{Stellar accretion}
\label{sec:stellacc}

Stars can be accreted by MBHs when they pass sufficiently close. The process has been studied in particular for cases that produce luminous emission (TDEs), but in principle stars can be swallowed whole, without producing radiation. This is the case for very massive MBHs only. We refer the reader to \cite{2021SSRv..217...40R} for a detailed review of the physical process, and we focus here only on their role in MBH growth. 

\citet{2017NatAs...1E.147A} were the first to realize that stellar accretion can play an important role in the growth of low-mass BHs, being able to source about $10^5-10^6 \msun$ over the Hubble time. The contribution to MBH growth can be estimated starting from the rate of stellar accretion/disruption. From the galaxy stellar density profile and the MBH mass one can derive expected rates \citep{2004ApJ...600..149W,2016MNRAS.455..859S}, which also depend on the presence or absence of nuclear star clusters \citep{2020MNRAS.497.2276P,2025ApJ...988...29H}. The rate peaks at about $10^{-3}-10^{-4}$ for MBHs with nuclear star clusters, and is lower by at least one order of magnitude otherwise, with significant scatter — a variety of galaxies and MBHs give rise to different rates at fixed galaxy or MBH mass. 

Following \citet{2017NatAs...1E.147A}, additional studies have been developed to assess the relevance of stellar accretion during MBH evolution in galaxies. \citet{2021MNRAS.500.3944P} included stellar accretion in a cosmological simulation run to $z=6$, finding that it contributes as much as gas accretion up until about $10^5 \msun$  of growth, making it particularly relevant when SNae can suppress gas accretion. \citet{2024A&A...689A.204P} implement stellar accretion in a semi-analytical model, run to $z=0$, and find that it contributes about $10^{-2}$ to the mass of $10^6 \msun$, decreasing to $10^{-4}$ for $10^8 \msun$.

\subsection{The Eddington limit}
\label{sec:Eddlim}

The Eddington limit applies to spherical sources of radiation embedded in hydrogen\footnote{The presence of metals and dust modifies the cross section, and therefore modulates the actual value of the Eddington luminosity. See \citet{2008MNRAS.385L..43F} for spherical symmetry and \citet{2020ApJ...900..174V}  for discs. See also \citet{2024A&A...684A.207F} for an application to galaxies – in spherical symmetry – and \citet{2025A&A...695A..33V} for an application to AGN/LRDs – also in spherical symmetry.}. When the outward radiation pressure exceeds to the inward gravitational force, radiation pushes away the gas. For stars this causes winds to unbind the outermost layers, for MBHs this would mean that further accretion is halted. In fact, for MBHs this is not the complete story. \cite{1979MNRAS.187..237B} already showed that even in spherical geometry radiation can be trapped, meaning that the time for photons to escape the disk exceeds the timescale gas to flow into the BH, and accretion can exceeed the Eddington limit -- although the luminosity will remain close to the Eddington luminosity. Furthermore, most BHs accrete from accretion discs, therefore a spherical configuration would not apply. In the case of accretion discs, the expectation is that the geometry and thermodynamics of the disc change as a function of the Eddington ratio, with thin, radiatively efficient discs below, but close to, the Eddington luminosity, and thick, radiatively inefficient discs at highly sub-Eddington and super-Eddington value \citep[see][for a review]{2015ASSL..414...45T}. Even in the case of discs, at super-Eddington rates radiation is expected to be trapped, leading to a $\sim$ logarithmic dependence of the luminosity on the accretion rate: highly super-Eddington accretion does not imply highly super-Eddington luminosities \citep[but see][]{2019ApJ...880...67J}. 

General-relativistic simulations show that super-Eddington accretion can create powerful jets \citep[][and references therein]{2016MNRAS.456.3915S}. When these jets are included in galaxy-scale or cosmological simulations the super-Eddington duty cycle can be low \citep{2019MNRAS.486.3892R,2023A&A...670A.180M}, but with sufficiently deep potential wells, low spins and non-extreme magnetic fields super-Eddington feedback is less damaging \citep{2023A&A...669A.143M,2024A&A...686A.256L,2025MNRAS.537.2559H}, and short phases of super-Eddington accretion can in principle grow the MBH more than prolonged phases of Eddington-limited accretion. Furthermore, growth is steadier if no jets are produced during super-Eddington
phases \citep{2016MNRAS.456.2993L,2020MNRAS.497..302T,2023MNRAS.519.1837S}. 

An important caveat is mass loss within the accretion disc: winds can limit the mass accreted onto the MBH to a small fraction of the accretion rate at the edge of the disc. The loss can be such that the MBH growth rate never exceeds the Eddington rate \citep{2012MNRAS.420.2912B,2025MNRAS.540.2820F}!

\subsection{MBH growth at high redshift}
\label{sec:highz}

\begin{figure}
\includegraphics[width=\textwidth]{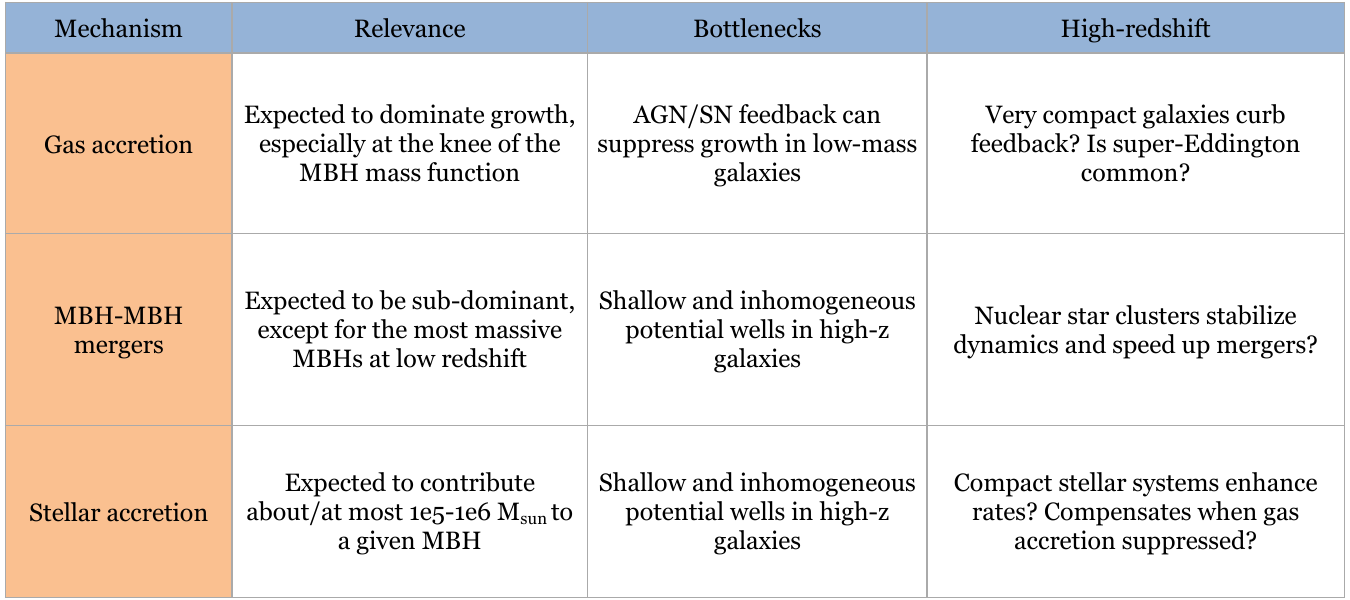}
\caption{Schematic view of MBH formation models and open questions.}
\label{fig:SchemeGrowth}       
\end{figure}

Given the MBH formation models described in Section~\ref{sec:form}, and the growth mechanisms outlined in this section, we can now focus on the growth of seeds in high-z galaxies. Fig.~\ref{fig:SchemeGrowth} provides a schematic summary.

The rare heaviest seeds formed in sites of strong UV radiation initially sit in the center of an almost spherical gas distribution, conducive to accretion and growth, but they are expected to be in satellites. Direct simulations of this environment \citep{2021MNRAS.502..700C} find that radiative feedback from the MBH itself suppresses the accretion rate (no kinetic/thermal feedback is included) and intense SN activity further injects a large amount of energy into the gas, causing supersonic turbulence, leading to sub-Eddington rates after formation. Furthermore, when the satellite merges with the main halo, the MBH accelerates leading to large relative relative to the gas, which causes a drop in the accretion rate. Accretion somewhat recovers after this, but the MBH is still wandering within the galactic potential at the end of the simulation.  With the same UV level as \citet{2021MNRAS.502..700C}, but included as a background, therefore without inclusion of dynamical effects, \citet{2019MNRAS.486.3892R} implement super-Eddington accretion, finding that the strong jets completely curtail the first growth spurt. More optimistic results are found if the MBH is placed (directly) in the center of a central halo – even in the presence of feedback from SNae and X-ray radiation (no UV radiation, kinetic or thermal feedback are included) from the MBH \citep{2020MNRAS.497.3761L}. 

Light PopIII seeds are scattered in a morphologically complex halo with few pockets of dense gas. \citep{2018MNRAS.480.3762S} find that the large majority of them is unable to grow because of random motions in shallow and irregular potential wells with turbulent gas, SN explosions and MBH feedback. Only about 0.1-1\% is found to be able to grow to $>10^4 \msun$ when MBH feedback is not very efficient and gas densities are so high that potential well deepens before SNe explode \citep{2024A&A...691A..24S}  or SNe push gas towards the BHs \citep{2024OJAp....7E.107M}.

For seeds forming in dense clusters, direct simulations of their evolution are still limited, mainly because forming massive, dense clusters requires cosmological simulations with very high-resolution, which therefore cannot be run for long times. Some simulations do resolve dense stellar and even nuclear clusters directly \citep{2025A&A...698A.207C,2025arXiv250308779G} or via re-mapping \citep{2023MNRAS.519.1366G}, thus we hope that this will happen soon. In the meanwhile, \citet{2023PhRvD.108h3012K}, for instance, find significant growth over time in isolated nuclear star clusters, which also gives good hope. In principle, MBHs embedded in a dense potential well should be able to grow more easily because feedback effects and erratic dynamics should be reduced \citep{2017MNRAS.469..295B}.  
Beyond gas accretion, as mentioned in Section~\ref{sec:mergers} MBH mergers appear challenging in shallow potential wells, and also in this case MBHs secluded in nuclear star clusters should have merge more easily. Stellar accretion appears to help in the early phases where gas accretion is hampered \citep{2021MNRAS.500.3944P}, and the combination of accretion of stars and mergers with stellar BHs, plus gas accretion can help MBHs grow fast \citep{2024arXiv241215334K}. 

\begin{overview}{Overview}
To summarize, gas accretion is expected to be the main growth channel for most MBHs in the Universe, but not necessarily at the very low and very high mass ends. In simulations, AGN feedback regulates both star formation and MBH growth, and SNe can limit MBH growth in low-mass galaxies. MBH mergers are also found to be challenging, except if MBHs are embedded in dense star clusters. Stellar accretion is generally subdominant, but can help when gas accretion is inefficient, provided -- again --  that MBHs are in dense stellar systems. Super-Eddington accretion in principle can help MBHs grow faster, but the strength of feedback limiting the duration, and the impact of mass loss are caveats that have to be fully explored. Growing MBHs appear therefore to be challenging -- and the inability to make them grow in simulations is challenged if the AGN population in low-mass galaxies detected by JWST is as large as currently considered. 
\end{overview}

\section{Modeling MBH Evolution}
\label{sec:models}
\begin{figure}
\includegraphics[width=\textwidth]{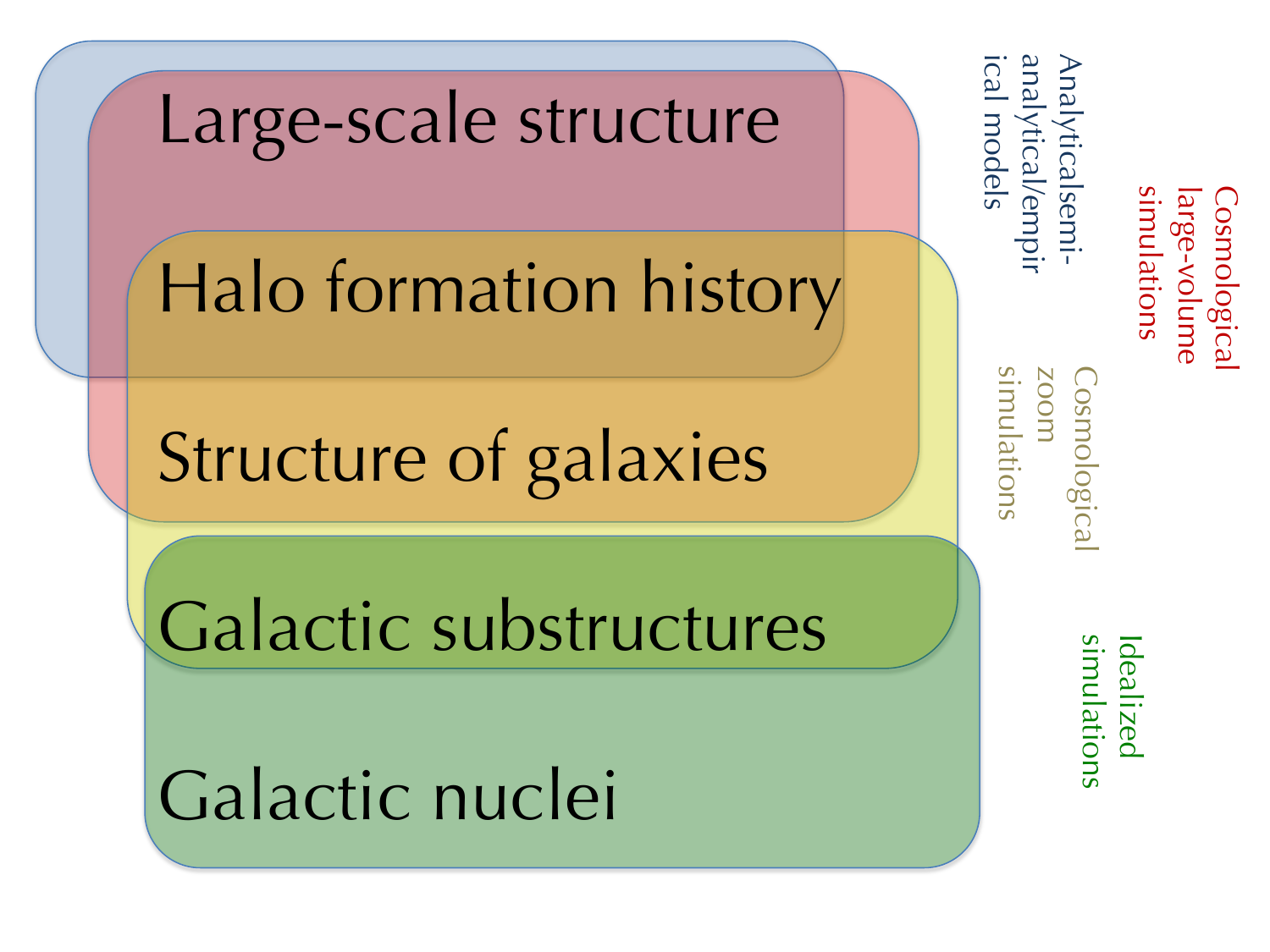}
\caption{Schematic view of the intrinsic range of various modeling approaches. Analytical and semi-analytical models can extend applicability outside this range, to smaller scales, with analytical recipes, while simulations can achieve the same through sub-grid physics. Figure concept: Michael Tremmel.}
\label{fig:SchemeModels}       
\end{figure}

To model MBH evolution in the cosmological context, four techniques are generally used: analytical models, empirical models, semi-analytical models, and cosmological simulations. To these, we can add idealized simulations of accretion discs, galaxy nuclei, individual galaxies, merging galaxies. What type of technique is best to use depends on the questions being asked (Fig.~\ref{fig:SchemeModels}).

Analytical models usually adopt an approach based on the \citet{1974ApJ...187..425P} formalism and its improvements \citep[e.g.,][]{2001MNRAS.323....1S,2008ApJ...688..709T}, where MBHs are associated to dark matter halo properties. They are clean and easily reproducible, but they do not allow to connect directly objects over time. 
Semi-analytical models also populate dark matter halos with MBHs (and galaxies), but they use dark matter halo merger trees obtained via the extended Press \& Schechter formalism \citep{1993MNRAS.262..627L} or from dark matter-only cosmological simulations. This permits to follow the history of individual halos. Semi-analytical models are fast, cover large parameter space, and give a good physical intuition of the general astrophysics. However, they lack spatial information, and can only use simplified analytical functions, therefore one looses control on non-analytical processes (those that cannot be described by well behaved mathematical functions, e.g., galaxy mergers). Empirical models populate DM halos with galaxies and MBHs, producing an ensemble population information compared to (many!) observables to find how the overall population evolves. A subset of empirical models are based on the continuity equation approach \citep{1971ApJ...170..223C}. Cosmological hydrodynamical simulations naturally include spatial information and can reach a high level of complexity, but they have  high computational costs and still require to model ``sub-grid'' physics that cannot be resolved. 

Personally, I first started working with semi-analytical models, and then moved to hydrodynamical simulations and empirical models. For the most part simulations have confirmed the results of semi-analytical models in terms of population statistics. What have I learned from simulations that I had not learned from semi-analytical models? Mainly things related to ``messy'' conditions or driven by environment, for instance the importance of the central galaxy density on the pairing fraction of MBHs \citep{2009ApJ...696L..89C},
the messiness of high-z galaxies questioning simple dynamical friction timescales \citep{2019MNRAS.486..101P}, SNae messing up gas near
MBHs \citep{2015MNRAS.452.1502D}, cosmological tidal fields influencing MBH growth \citep{2017MNRAS.467.4243D}. A typical trend nowadays is in fact to implement in semi-analytical models recipes based on the results of hydrodynamical simulations. What have I learned from empirical that I had not learned from semi-analytical models and simulations? That data-driven models can help understand what works and does not work in semi-analytical models and simulations, that matching a variety of observational constraints is really hard \citep{2023MNRAS.518.2123Z}, that accretion in dwarf galaxies is more consistent with low duty cycles than low mean accretion rates \citep{2024MNRAS.529.2777Z}, that modeling the AGN population in radio is very complex \citep{2021A&A...650A.127T}. 

In the following I will focus on cosmological simulations, and I refer the reader to, e.g.,  \citet{1998ApJ...503..505H,2002ApJ...581..886W} for analytical models, e.g., \citet{2000MNRAS.311..576K,2003ApJ...582..559V,2006MNRAS.365...11C,2006MNRAS.373.1173F,2008MNRAS.391..481S,2011MNRAS.410...53F,2012MNRAS.423.2533B,2019MNRAS.486.2336D,2020MNRAS.495.4681I,2022MNRAS.511..616T} for semi-analytical models and e.g.,  \citet{2009ApJ...690...20S,2021A&A...650A.127T,2023MNRAS.518.2123Z} for empirical models; where I have tried to cite papers written by different groups (typically the first of the series: many papers usually follow!). I will note in particular that since semi-analytical models are not limited (as much) by resolution, they are able to implement various types of MBH formation mechanisms, down to light seeds. 

\subsection{Cosmological simulations}
\label{sec:CosmoSims}

Cosmological simulations are thoroughly reviewed in \citet{2020NatRP...2...42V}. Regarding MBHs in simulations, \citet{2021MNRAS.503.1940H,2022MNRAS.509.3015H} discuss and compare several different implementations.  I refer the reader to these papers for a comprehensive background, and I will only summarize the main features. In general, I will try to reference in the following the papers where a given approach has been presented for the first time.

\begin{figure}
\includegraphics[width=\textwidth]{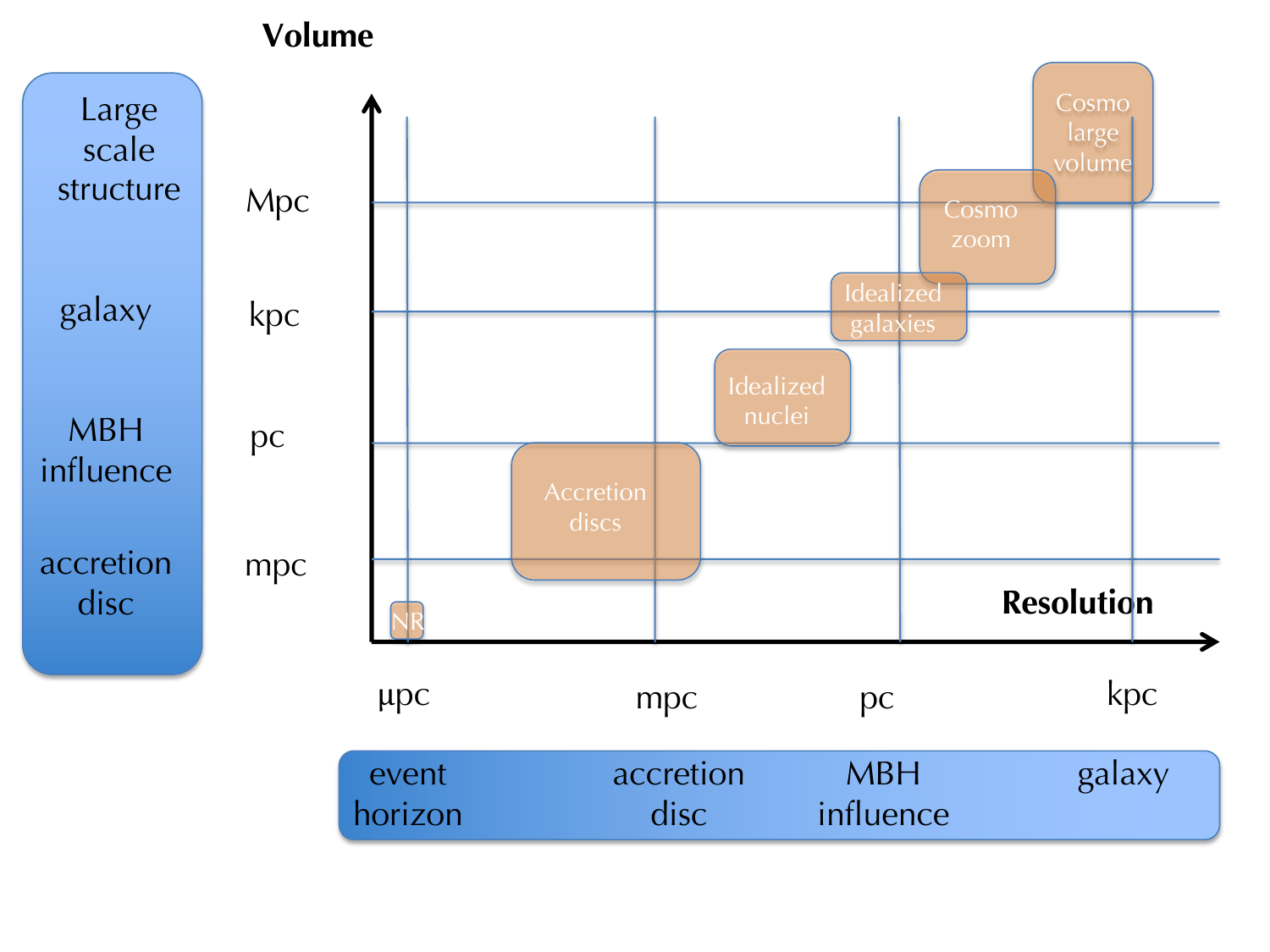}
\caption{Schematic view of volume vs resolution for different type of simulations, along with the physical scales probed. Besides the cosmological simulations discussed here the scheme includes also idealized galaxies (including galaxy mergers), accretion disc simulations and numerical relativity (`NR' in the scheme).}
\label{fig:SchemeSims}       
\end{figure}

Simulations are like observational surveys: you can have either large and shallow (large volume/many objects/low resolution/massive galaxies) or small and deep (small volume/few objects/high resolution/dwarf galaxies). Large volume boxes simulate a given volume of the Universe with the same resolution everywhere (still with local refinement). The large volume, which probes the large scale structure and provides statistics on a large number of relatively massive galaxies, limits the resolution. Normally spatial resolution is $\sim$ kpc, particle mass resolution $\sim 10^6 \msun$ (and one needs at least $\sim$~50 particles to define a galaxy). Running a cosmological simulation with these specs to $z=0$ takes tens of millions of CPU hours: several months to years of real time! In zoomed simulations a specific area in a box is resampled and re-simulated at higher mass and spatial resolution. Since only a small volume is resimulated the computational cost is lower, but the number of halos in the volume is smaller: we loose statistics on the global population. 

In all cases, the typical resolution of simulations does not allow to resolve all necessary scales (see the discussion in Section~\ref{sec:basics} and Fig.~\ref{fig:SchemeSims}). Simulations account correctly for gravity and hydrodynamics on resolved scales, but for smaller scales and unresolved processes one needs to implement ``sub-grid'' prescriptions, akin to ``recipes'' in semi-analytical models. Only in one cosmological simulation zooming onto a single MBH for a few thousands of years, resolution of all relevant scales, down to the accretion disc and the MBH horizon has been performed \citep{2024OJAp....7E..18H,2025OJAp....8E..48H}: an incredible feat!  

\subsubsection{MBH formation}
\label{sec:simsMBHform}

MBHs are typically initialized using two approaches: one is to place them in halos (or galaxies) above a certain threshold using an on-the-fly halo finder \citep{2008ApJ...676...33D}. The other is to place MBHs in high gas- and stellar-
density regions, mimicking MBH formation mechanisms \citep[e.g.,][]{2017MNRAS.470.1121T,2017MNRAS.468.3935H}. As discussed in Section~\ref{sec:form}, models predict initial masses $\sim 10^2-10^5 \msun$, but normally the choice is limited by mass resolution: to avoid numerical noise the MBH mass should be a factor of $\sim 10$ larger than the mass of other particles \citep{2015MNRAS.451.1868T}. A way to circumvent this limitation is to assume that MBHs are born in dense stellar clusters, and decouple the MBH mass for accretion (the ``real'' mass) and the dynamical mass (MBH+cluster). This allows to decrease the ``actual'' MBH mass while at the same time to stabilize its dynamics \citep{2017MNRAS.469..295B}. 

\subsubsection{Gas accretion and AGN feedback}
\label{sec:simsgasacc}

There are three main approaches to mimicking gas accretion onto MBHs. The first adopted, and still the most common, relies on modifications of the Bondi-Hoyle-Littleton \citep{1939PCPS...35..405H,1944MNRAS.104..273B} formalism \citep{2008ApJ...676...33D}:

\begin{equation}
    \dot{M}_{\rm BH}=\alpha 4 \pi G \rho \frac{\mbh^2}{(c_s^2+v_{\rm rel}^2)^{3/2}},
\end{equation}

where $\rho$ is the gas density and $v_{\rm rel}$ is the relative velocity between MBH and gas.  $\rho$, $c_s$ and $v_{\rm rel}$, or $\dot{M}_{\rm BH}$ are averaged over the MBH kernel. $\alpha$ is a fudge factor required at low resolution to
capture high accretion rates due to unresolved large density contrasts, and is often related to the gas density threshold for star formation \citep{2009MNRAS.398...53B}. The second is to assume that gas inflows are driven by torques \citep{2017MNRAS.464.2840A}, the third, possible only at extremely high resolution, is to calculate directly the mass flux onto MBHs \citep{2019MNRAS.486.3892R,2021ApJ...917...53A}. 

Further, many simulations cap the accretion rate to the Eddington luminosity, generally assuming a fixed radiative efficiency \citep{2008ApJ...676...33D}, although recently super-Eddington physics has been implemented in some simulations \citep{2019MNRAS.486.3892R,2023A&A...670A.180M,2024A&A...686A.256L,2025MNRAS.537.2559H}. 

AGN feedback is generally implemented as injection of thermal or kinetic energy: a fraction $\epsilon_f$ of the accretion or radiative luminosity is re-injected in the MBH's surroundings. In the case of thermal input the gas temperature is increases by distributing specific energy in a small sphere near the MBH \citep{2008ApJ...676...33D}. For winds, outflows are injected  with high
velocity (typically $10^4 \kms$) close to the MBH \citep{2012ApJ...754..125C}, while for jets the outflows are collimated, e.g., injected with cylindrical geometry \citep{2012MNRAS.420.2662D}. The feedback efficiency $\epsilon_f$ is generally calibrated by fitting for the MBH scaling relations, and typically feedback is stronger at higher resolution because more energy is injected per unit volume: one needs therefore to decrease the efficiency as resolution increases \citep{2017MNRAS.467.3475N,2019MNRAS.488.4004L}.

\subsubsection{MBH dynamics and mergers}
\label{sec:simsdyn}

Early simulations related the MBH position to the center of the host, repositioning the MBH at every time step to the minimum of the local potential \citep{2008ApJ...676...33D,2009ApJ...690..802J}. More recently, the ``missing'' dynamical friction caused by limited resolution has been added explicitly, for gas, stars and dark matter, or all species \citep{2013MNRAS.428.2885D,2015MNRAS.451.1868T,2019MNRAS.486..101P}. Even more recently, additional physics related to binary evolution has been accounted for. The first approach is to use a high-accuracy regularized integrator in a small spherical region with typical radius of $\sim 10$~pc centered on the MBH \citep{2023MNRAS.524.4062M}: this approach models accurately stellar hardening, three-MBH interactions and GW emission, but does not have the same accuracy for gas-driven processes. The second is to develop a sub-grid model that evolves the MBH dynamics accounting for dynamical friction,  stellar hardening, migration in a circumbinary disc and GW emission \citep{2024arXiv241007856L}. The latter two approaches allow for MBH mergers to occur when the separation is a few $R_g$, as in reality. Conversely,  in the previous approaches MBHs where merged using a proximity criterion in terms of N resolution elements \citep{2008ApJ...676...33D}, which correspond to hundreds of pc to kpc for most simulations, 3-5 orders of magnitude larger than separation at which MBHs actually merge. The proximity criterion can be supplemented by a dynamical criterion, e.g., the MBHs are bound \citep{2017MNRAS.470.1121T}. The caveat for the proximity (and proximity+dynamical) approach is that key orbital parameters such as mass ratio and eccentricity near the merger, necessary for waveform modeling, data analysis and electromagnetic signatures of MBH mergers cannot be  predicted. 

\subsubsection{MBH spins}
\label{sec:simsspin}

MBH spins have received less attention than their masses, but they have been included in cosmological simulations, accounting for spin evolution by accretion and MBH-MBH mergers \citep{2014MNRAS.440.2333D}. For MBH-MBH mergers the results from simulations of MBH mergers in numerical relativity are used, while for accretion the evolution considers the relative direction of spin and gas angular momentum. Angular momentum is transferred from disc to MBH, with spin up or down depending on disc/MBH mass ratio and alignment as well as on spin extraction in jets. 

\subsubsection{Other physical processes}
\label{sec:simsother}

Some simulations include additional physical properties, such as stellar accretion and TDEs \citep{2021MNRAS.500.3944P} or cosmic rays from AGN \citep{2008MNRAS.387.1403S,2025A&A...696A..58V}.

\begin{overview}{Overview}
Modeling MBH evolution in the cosmological context can be performed with different techniques, where the choice of the ``best'' technique depends on the questions one wants to investigate. The main challenge is the vast dynamical range of spatial (and time) scales involved. Typically, simulations need to model processes below resolution as ``sub-grid'' models, with different groups making different choices. 
\end{overview}

\section{Modeling MBH observables}
\label{sec:observables}

We normally detect MBHs when they emit electromagnetic or GW radiation, which corresponds to times when the MBH is either accreting gas or stars, or merging with another compact object. In the local Universe some quiescent MBHs have also been detected, via their dynamical imprint on the orbits of stars and gas in their environment: these are the MBHs that have allowed us to study their demography and scaling relations with the host galaxies. I will not touch on these quiescent MBHs and I refer the reader to \citet{2009ApJ...698..198G,2014SSRv..183..253P}. 

\subsection{AGN}
\label{sec:AGN}

The emission from AGN is determined by the properties of the accretion discs and their surroundings. It is considered that the structure of accretion discs depends on the Eddington ratio, from thick, radiatively inefficient discs at the low and high Eddington ratio ends (highly sub-Eddington and super-Eddington) to thin, radiatively efficient discs in the mildly sub-Eddington regime. The typical transitions occur around $\fedd\sim 10^{-3}-10^{-2}$ and $\fedd\sim 0.3-1$ \citep[e.g.,][and references therein]{2015ASSL..414...45T}.  At high-z the large majority of sources has high Eddington ratio, simply because they need to be intrinsically bright to be observed at large distances. I will therefore focus on the high Eddington ratio regime. 

\subsubsection{Thin disc emission}
\label{sec:thinkdisc}

For thin, radiatively efficient discs, one can develop analytical models assuming mass, angular momentum conservation and that energy is dissipated locally, one can obtain a temperature profile within the accretion disc: a multicolor black-body (\citealt{1973A&A....24..337S,1973blho.conf..343N}, see \citealt{2022agn..book..101L} for a review). For MBHs the peak temperature is in the rest-frame optical/UV: 
\begin{equation}
    T_{\rm in}=10^5 \fedd^{1/4} \left(\frac{\mbh}{10^8 \msun}\right)^{-1/4} \left(\frac{R}{R_g}\right)^{-3/4} \rm{K}
\end{equation}
increasing with decreasing mass and accretion rate (cf. accreting stellar mass BHs are called X-ray binaries because the bulk of the emission is in X-rays). Photons from the disc are assumed to reach an optically thin region (“corona”) where they are upscattered and generate the X-ray emission (power-law). 

\begin{figure}
\includegraphics[width=\textwidth]{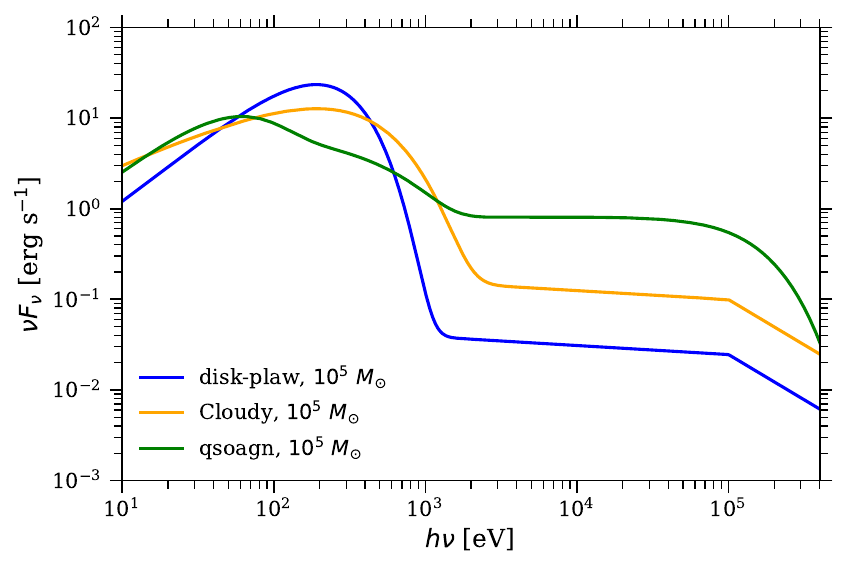}
\caption{Comparison of different SEDs obtained fixing the MBH mass and accretion rate ($\mbh=10^6 \msun$, \fedd=0.1). Adapted from \citet[][Chris Richardson, private communication]{2022ApJ...927..165R}.}
\label{fig:SEDcompare}       
\end{figure}

When modeling the SEDs from thin discs, there are typically three options. The first is fully consistent, first principle models, which consider the baseline disc emission and further include comptonization as well as reprocessing to estimate intrinsic SED \citep[e.g.,][available in XSPEC as ``QSOSED'']{2018MNRAS.480.1247K}. The second option is to use a combination of power-laws to mimic observed SED of quasars \citep[e.g.][]{2004MNRAS.351..169M,2007ApJ...654..731H}, an approach that is often included in SED fitting codes \citep[e.g.,][]{2020MNRAS.491..740Y,2024MNRAS.527.7217V}. The third option is to assume a functional form mimicking the results from first principle models and fix parameters based on observations \citep{2013RMxAA..49..137F,2016ApJ...833..266T,volonteriHighredshiftGalaxiesBlack2017}. The resulting SEDs for a given MBH can be significantly different depending on the approach chosen, as shown in Fig.~\ref{fig:SEDcompare}. 

\subsubsection{Super-Eddington disc emission}
\label{sec:sEdddisc}

As already mentioned, it is expected that super-Eddington AGN are characterized by radiation trapping: the time for photons to escape the inner part of the disc
becomes very long because of the high densities – the diffusion time gets longer
than the inflow time. Instead of leaving the disc and reach an observer
photons are advected into the BH \citep{1979MNRAS.187..237B}.  This limits the local emissivity to about the Eddington luminosity \citep[with a $\sim$ logarithmic correction, see][and references therein]{2022agn..book..101L}, while at the same time the disc puffs up and in slim disc models causes a flatter dependence of temperature on radius. 

A simple approach to mimic radiation trapping is to suppress photon emission inside the trapping radius, $\sim \fedd R_g$, with a truncation of the inner disc \citep{2017MNRAS.466.2131P,2020MNRAS.492.4058P} or a shift of the peak frequency \citep{2025A&A...695A..33V}, both having the effect of making the SED redder. 
A more complex, but more physical, approach accounts for effects on radiative processes in the disc \citep[][available in XSPEC as ``AGNSLIM'']{2019MNRAS.489..524K}.

\subsubsection{Obscuration and attenuation}
\label{sec:obsc}

In most cases what we see is not directly the emission from the accretion disc, because of the gas and dust along the line of sight, in particular the dense gas near the MBH and in the host galaxy.  The role of gas and dust in the vicinity of the MBH in shaping the AGN emission has been recognized long ago \citep{1995PASP..107..803U,1993ARA&A..31..473A}. When including obscuration and attenuation in models, typically two approaches have been adopted. The first is to extrapolate column density distributions \citep[e.g.,][]{2014ApJ...786..104U} or obscured/absorbed fraction from observations \citep[e.g.,][]{2014MNRAS.437.3550M}. The second is to estimate column density distributions distribution and/or obscured/absorbed fraction directly from cosmological simulations \citep[e.g.,][]{2025A&A...695A..33V,2019MNRAS.487..819T,2020MNRAS.495.2135N}. Since cosmological simulations generally don't resolve the sub-pc to pc scales where the densest gas resides, this can be added using, for instance, simple analytical models \citep{2023A&A...676A...2D}, or dedicated tools \citep[e.g.,][]{2008ApJ...685..160N,2016MNRAS.458.2288S}. 

\subsubsection{Galaxy + AGN models}
\label{sec:AGNgal}

Ideally, one would like to have the AGN SED along with the galaxy SED\footnote{There are several tools dedicated to synthetic observations for galaxies, without AGN. I will not comment on them here, and refer the reader to the Chapters by Richard Ellis and Rachel Somerville}. Some of the tools developed to fit galaxy or AGN SEDs can also be used to extract galaxy or AGN SEDs from models, as done for instance with BEAGLE in \citet{2017MNRAS.472.2468H}. A tool specifically dedicated to post-process simulation outputs to obtain synthetic observations is Synthesizer \citep{2025arXiv250615811R}, which handles both stellar populations and AGN, along with gas and dust. 

\begin{figure}
\includegraphics[width=\textwidth]{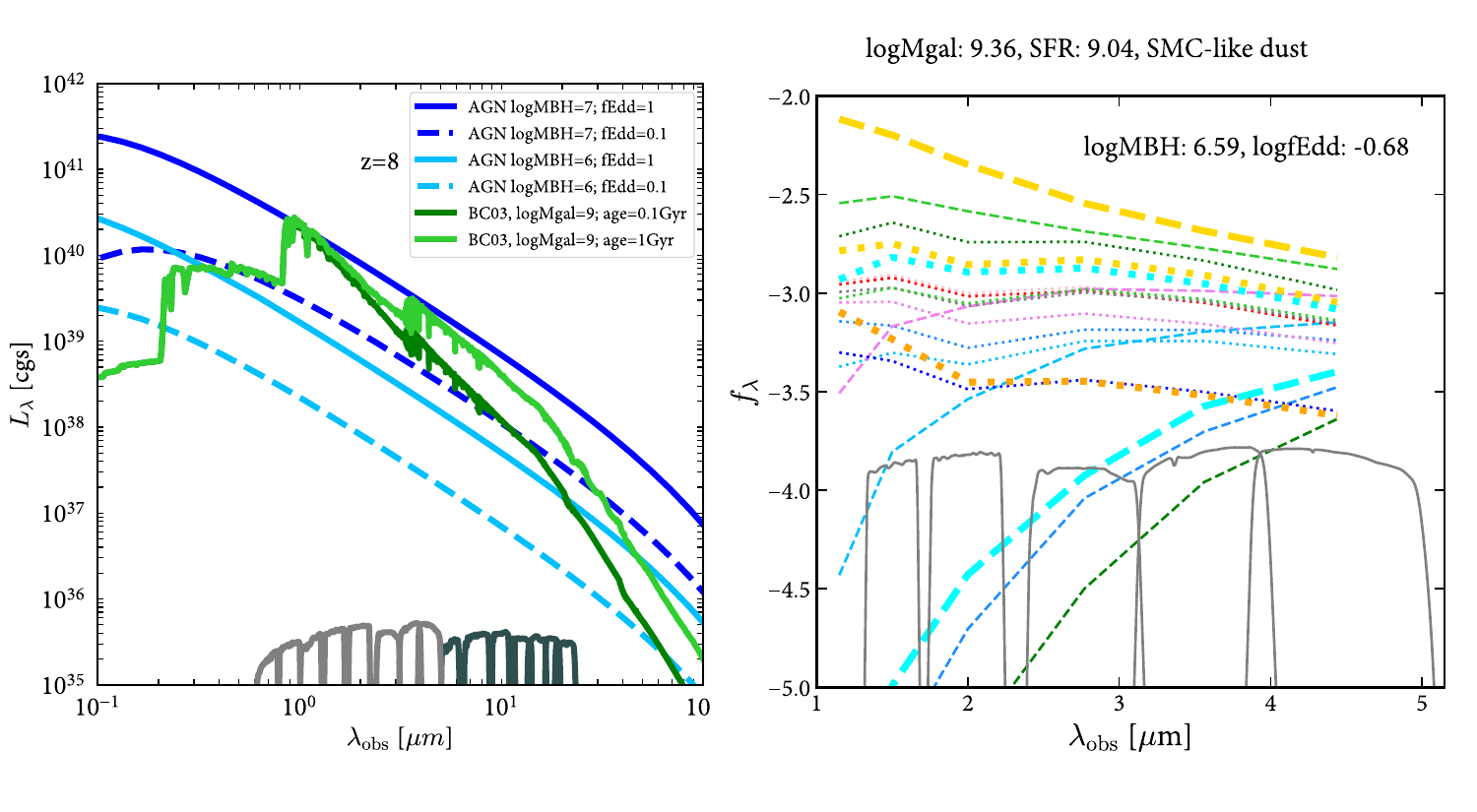}
\caption{Left: intrinsic (non-attenuated) example galaxy (green solid curves) + AGN SEDs (blue solid and dashed curves). Here the galaxy is assumed to have constant star formation rate. Galaxy masses and ages, as well as AGN masses and Eddington ratios are reported in the legend. NIRCAM and MIRI filters are reported at the bottom in gray. Right: example of SEDs for a galaxy extracted from the Obelisk simulation \citep{2021A&A...653A.154T}, with the addition of an AGN, observed over 12 different lines of sight. For each color the dashed curves report the emergent AGN flux, the dotted curves the emergent galaxy flux. Gaalxy and MBH properties are reported in the legend. NIRCAM filters are shown at the bottom in gray.}
\label{fig:SED_agngal}       
\end{figure}

As an example, I will consider the simple approach developed in \citet{volonteriHighredshiftGalaxiesBlack2017,2025A&A...695A..33V}. The AGN SED is obtained from the MBH mass and accretion rate using an empirical approach, aimed at reproducing the typical AGN SED, but allowing for a mass and mass ratio dependence \citep[see also][]{2022ApJ...929...21Z}. This is combined with a galaxy SED, obtained from simple stellar populations or from the star formation history of simulated galaxies. The left panel in Fig.~\ref{fig:SED_agngal} shows some examples, highlighting that AGN contribute significantly to the SED in the JWST bands for high-z sources only when the MBH mass is $>10^{-2}$ the galaxy mass (i.e., the MBH is overmassive with respect to the $z=0$ relation) and it has high \fedd. One can also appreciate that for young, star forming galaxies the stellar population is ``bluer'' than the AGN \citep[as found also in][]{2017ApJ...838..117N}. Such intrinsic SEDs can then be attenuated by considering the gas and dust distribution in simulated galaxies, for instance the Obelisk simulation \citep{2021A&A...653A.154T}. The right panel in Fig.~\ref{fig:SED_agngal} shows how different the emergent SED along 12 lines of sights can be. The AGN emission, in particular, is more strongly affected by the line of sight. This is because AGN attenuation is dominated by few dense gas clouds in the galaxy center. The same system can appear galaxy-dominated or AGN dominated depending on \emph{both} intrinsic properties (galaxy mass, star formation rate, \mbh\ and \fedd) and the distribution of absorbers. 

Applying this model to simulated galaxies at $z=6$ \citep{2025A&A...695A..33V}, it found that a color selection mimicking that of LRDs selects AGN with intermediate dust column density/attenuation: too little attenuation makes sources too blue, too much attenuation makes them too faint or too sub-dominant with respect to the host galaxy. Incidentally, this suggests that the escape fraction from AGN is not high, at least for UV photons. Detectable AGN (above an AGN luminosity threshold to detect broad lines, and above a combined AGN+galaxy magnitude to enter JWST's photometry) have enhanced high Eddington ratios and/or high \mbh-\mstar\ ratios, and color-selected AGN exacerbate these trends.

\subsubsection{Radio emission}
\label{sec:radio}

Models to predict the radio emission from modeled AGN based on intrinsic properties available in population models (mass, accretion rate and spin) are rarer, mainly because they are not sufficient to fully determine radio properties, which depend also on magnetic field strength and the intrisic properties of jets. I refer the reader to \citet{2018MNRAS.475.3493B}, who in their Appendix provide guidelines on how to calculate radio emission in magneto-hydrodynamical simulations, and \citet{2018MNRAS.475.2768H}, who develops a semi-analytical model based on simulation results. 

Simplified approaches adopt either scaling relations based on jet models \citep{2001ApJ...548L...9M}, and the emission at a given frequency can be calculated assuming, e.g., a power-law spectrum \citep{2023A&A...676A...2D}, or extrapolations of the Fundamental Plane of BH activity, linking X-ray and radio luminosity to BH mass \citep{2003MNRAS.345.1057M,2004A&A...414..895F,2009ApJ...706..404G,2012MNRAS.419..267P}.  A priori  the Fundamental Plane is valid only for highly sub-Eddington BHs, and global applicability is disputed \citep{2022MNRAS.516.6123G}.

\subsection{GWs}
\label{sec:GWs}

The GW frequency emitted by merging compact objects can be approximated as $f\sim c^3/(2 \pi G \mbh)\sim 10^4 \rm{Hz} \, \mbh/\msun$, by simply assuming that the speed cannot exceed the speed of light. For stellar-mass BHs, this falls in the LIGO-Virgo-KAGRA frequency, while for MBHs with mass $\sim 10^4-10^6 \msun$ the $\sim$~mHz frequency range will be covered by LISA \citep[$\sim 10^{-4}-10^{-2}$~Hz,][]{2017arXiv170200786A} and masses $\sim 10^8-10^{10} \msun$ pertain to the $\sim$~nHz frequency range probed by Pulsar Timing Arrays \citep[PTAs, which probe in fact the GW inspiral, rather than the actual merger;][]{1983ApJ...265L..39H}. 

To model LISA GW observables \citep{2019CQGra..36j5011R,2021arXiv210801167B}, the simplest option is to use the instrument response averaged over sky position, polarization and source inclination along with a waveform model, where care is needed for mass ratios $<1:20$, where current waveforms are less accurate, and to consider the appropriate spin limits for magnitude and direction in relation to the chosen waveforms. More refined approaches can be used to recover the measured source parameters (``parameter estimation''). In this case one has to use non sky-averaged instrument response, and randomize (or choose, if there are reasons for it,) sky position, polarization and source inclination. Parameter estimation via Fisher matrix is a fast option, accurate for most purposes \citep[e.g.,][]{2015PhRvD..91j4001P}, but Bayesian approaches, especially including higher harmonics, helps reducing degeneracies \citep[e.g,][]{2021PhRvD.103h3011M} at the cost of longer computational times. For applications to MBH populations, see, for instance \citet{2020PhRvD.102h4056M} and \citet{2022PhRvD.106j3017M}.

\begin{figure}
\includegraphics[width=\textwidth]{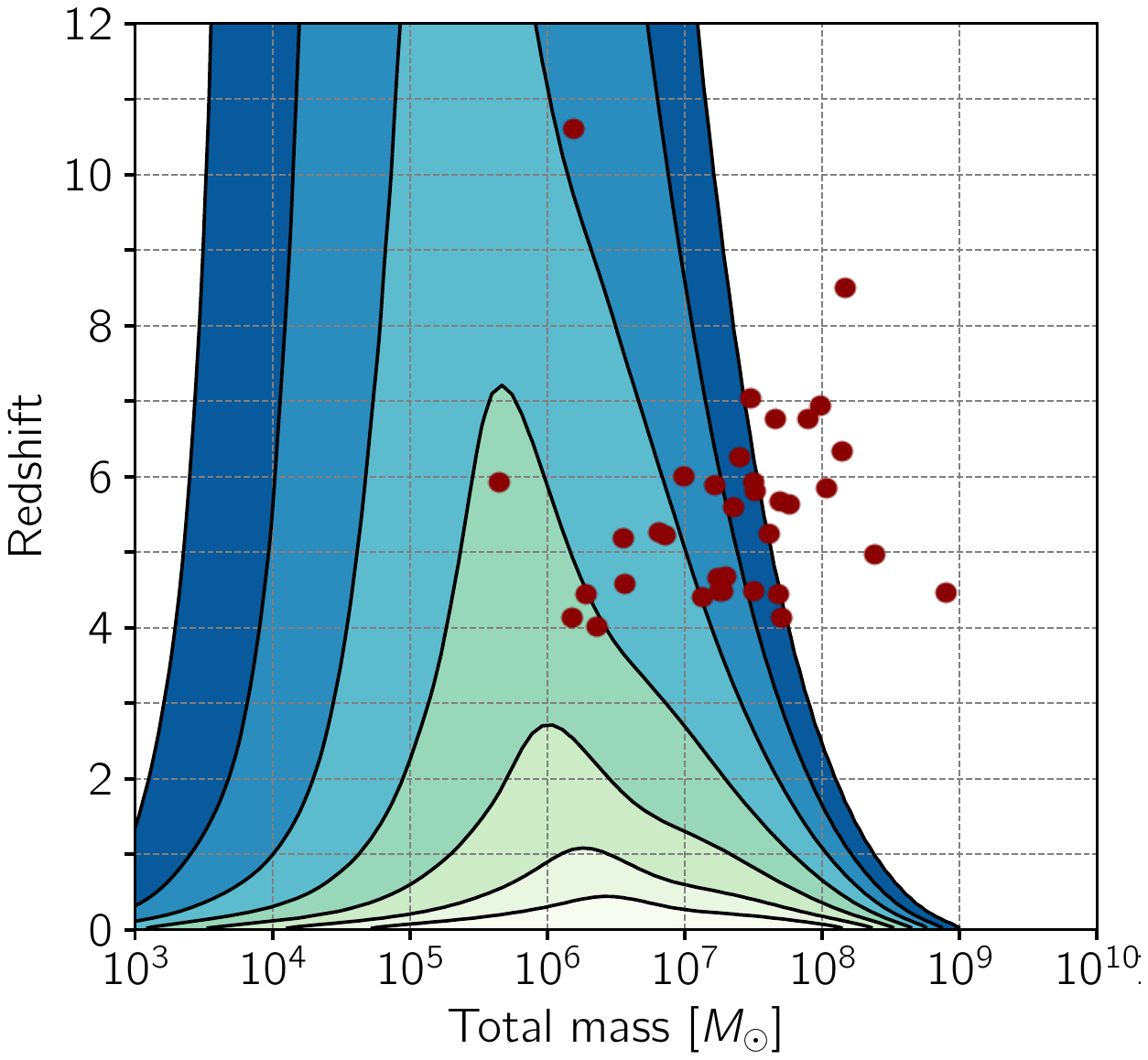}
\caption{LISA's reach in the \mbh-redshift plane (blue contours), compared to JWST's AGN (red dots). LISA's binaries have mass ratio 0.5 and aligned spins of magnitude 0.2, and the contours, extracted from \citet{2024arXiv240207571C}, show signal-to-noise ratios from 10 (dark blue) to $10^4$ (light yellow).}
\label{fig:LISA_JWST}       
\end{figure}

As shown in Fig.~\ref{fig:LISA_JWST}, LISA can detect merging MBHs at earlier cosmic times and with lighter MBHs than otherwise possible, including the region where heavy-ish MBH seeds are expected to reside (if they are able to merge!) With gravitational waves, one gets exquisite measurement of \mbh, at sub-percent level, up to few percent for sources with low signal to noise ratio \citep{2024arXiv240207571C}. This means that we will get totally complementary, and accurate, measurements of MBH masses, although one has to account for biases in the merging population with respect to the parent full MBH population \citep{2023A&A...673A.120D}. The luminosity distance is also generally measured with similar uncertainties, which means that if one assumes a cosmology the redshift can be inferred. Conversely, if an electromagnetic counterpart of the MBHs or the host galaxies is detected, redshift can be measured independently and constrain cosmological parameters \citep{1986Natur.323..310S,2016JCAP...04..002T,2025PhRvD.111h3043M}. For some events with low signal to noise ratio, distances and masses can be highly overestimated, which may be interpreted erroneously as evidence for early seeds or primordial BHs. These high deviations from the true values are however accompanied by large uncertainties, which offer a way to flag them as outliers \citep{2023A&A...676A...2D}. A negative note is that for high-z merging MBHs the sky localization is limited, typically $>>10 \, \rm{deg}^2$ up to the whole sky (!), meaning that finding the electromagnetic counterpart and the host galaxy will be challenging. Some approaches have been delineated in \citet{2016JCAP...04..002T,2022PhRvD.106j3017M,2023A&A...676A...2D,2023A&A...677A.123I}. 

An important point of note is that when modeling the GW signal from merging MBHs, some parameters are randomized (e.g., inclination, position in the sky), but for real sources this will not be the case! If the merger rate is low, then luck in the extrinsic parameters that contribute to the signal-to-noise ratio will be important. Currently the merger rates from models have large variations, from a few to a few thousands per year, in dependence of numerical aspects and astrophysical choices \citep[e.g., seed properties, efficiency of orbital decay,][]{2023LRR....26....2A}

As a low-redshift aside, in 2023 many groups around the world have reported the detection of GWs at nHz frequency and considered the inferences on the MBH population that could produce this signal  \citep{2023ApJ...952L..37A,2024A&A...685A..94E,2023ApJ...951L...6R,2023RAA....23g5024X}. Given the frequency and sensitivity of Pulsar Timing Arrays,  $\mbh \sim 10^8-10^{10} \msun$ and $z\lesssim 2$. The bottom line is that the signal, at the currently suggested levels, implies a more abundant population of MBHs than expected. I will note here that previous upper limit to the GW background at nHz frequency were \emph{lower} than the signal currently measured. Indeed, for many years theoretical models have been trying to suppress the signal \citep{2013Sci...342..334S}, very similarly to what has happened with the JWST detections of bright and numerous AGN. My lesson learned from this is to avoid modifying theoretical models to accommodate observational results until one is really really really sure that they are correct.

\subsection{TDEs}
\label{sec:TDEs}

I refer the reader to \citet{2021ARA&A..59...21G} for a comprehensive review of observational properties of TDEs. TDEs have a characteristic ligh curve \citep{1988Natur.333..523R}, decaying over time as $t^{-5/3}$ \citep{1989IAUS..136..543P}, with the bright phase lasting about 1 yr rest-frame, which has to be multiplied by $(1+z)$ to obtain the duration in the observer's frame. The early phase can be at super-Eddington levels \citep{2009MNRAS.400.2070S} at least for low-mass ($\lesssim 3\times 10^7 \msun$). Some TDEs span the full EM
spectrum, while some TDEs are mainly bright in UV/optical, while others are bright in X-rays. This behavior has been ascribed to superEddington accretion with varying lines of sight \citep[][more recently revived by the X-ray weakness of high-z AGN]{2018ApJ...859L..20D}. 

Their maximum luminosity is expected to be $\sim 10^{44}-10^{45} \ergs$ in dependence of \mbh\ and stellar mass/type \citep{2020ApJ...904...98R}, which makes it difficult to detect them at very high redshift. The current record holder is at $z\sim 5$ \citep{2025arXiv250413248K} and suggests that  deep surveys with JWST and Roman can find TDEs at $z>4$ and contribute to probing the MBH population at early cosmic times \citep{2024ApJ...966..164I}. The Legacy Survey of Space and Time (LSST), and Euclid can cover the redshift range up to $z\sim 4$. Some, or many, of the ``bizarre'' high-z AGN and LRDs may be TDEs \citep{2025ApJ...984L..55B}.

\begin{overview}{Overview}
Theoretical SED models for AGN work reasonably well from optical to X-ray, while modeling radio emission based on the quantities known in models of MBH cosmic evolution is more complex. 
Modeling GWs for astrophysical purposes is a well developed field, but care must be placed on the dependence on parameters such as  inclination, sky position etc. Normally these are
averaged over but for individual (real!) sources it will be different. 
TDEs are a new probe with great potential. 
\end{overview}

\section{Before and after JWST}
\label{sec:before_after}

After presenting an overview on the theoretical models on MBH evolution in the first billion years of the Universe, I want to circle back to Section~\ref{sec:JWST}, which can now be viewed by the reader with more information. I will also touch upon results and ideas that have  surfaced after the school took place. 

JWST has uncovered a large population of high-z AGN, which, taken at face value and estimating their properties with the same approaches used for low-redshift AGN, are inconsistent with extrapolation of pre-JWST observations. New analyses of the Hubble Ultra Deep Field using variability \citep{2025arXiv250117675C}, however, find very high number densities of AGN at all redshifts, and also new X-ray analyses find an enhanced number density of AGN at high-z \citep{2025arXiv250616145B}.  Have we really missed all these AGN? How much of this was a sociological bias? From personal experience, suggesting that high-z galaxies should host AGN, because $z>6$ bright quasars cannot be formed without implying a population of fainter sources, was considered silly.

Something that, as mentioned, would be useful is to assess whether  the bright end of the AGN LF, in the quasar regime, also been underestimated pre-JWST. The $z=0$ \mbh-$M_{\rm star}$ relation with its intrinsic scatter can easily accommodate the JWST AGN with an active fraction
of $\sim$25\% \citep{volonteriHighredshiftGalaxiesBlack2017,greeneUNCOVERSpectroscopyConfirms2024}, but there would be more AGN also at the bright end. This would be exacerbated if the \mbh-$M_{\rm star}$ relation evolves with redshift and MBHs, as a population, are overmassive with respect to today \citep{2023ApJ...957L...3P}. An alternative would be that the Eddington ratio has a strong dependence on galaxy or MBH mass, with most of the MBHs at the high-mass end being quiescent -- for instance having already reached self-regulation \citep{1998A&A...331L...1S,2003ApJ...596L..27K}. This would be somewhat in disagreement with the anti-hierarchical picture \citep{2004MNRAS.353.1035M}, according to which the most massive MBHs are most active at high-redshift. 

A couple of sources, UHZ1 \citep{2024NatAs...8..126B} and GHZ9 \citep{2024ApJ...965L..21K} are reported with such high masses $\sim 10^8 \msun$ and redshifts ($z\sim 11$), and implying such a high number density, that challenge theoretical models based on gas/stellar-origin seeds, and hinting at primordial BHs \citep{2024A&A...690A.182D}. A similar hint is suggested by the extremely low metallicity of a $z\sim 7$ LRD in comparison to models, coupled to the high inferred MBH mass \citep{2025arXiv250522567M,2025arXiv250821748J}. 

Regarding the overmassiveness of high-z MBHs, in other words an evolution with redshift of the \mbh-\mstar\ relation, while I stress that care must be exercised in how to interpret and compare samples, there are arguments to suggest that it would be a physical aspect of galaxy evolution: if the driving correlation is between \mbh\ and velocity dispersion, the increased compactness of halos (and galaxies) at high-z would imply that the ratio between  \mbh\ and \mstar\ should increase \citep[see][for this argument in relation to halo mass and circular velocity]{2003ApJ...595..614W}.

Considering the material presented in Section~\ref{sec:highz}, it seems difficult, theoretically, to grow MBH seeds and grow MBHs in low-mass galaxies, generically. A first point to note is that semi-analytical models are more able to obtain growth \citep{2022MNRAS.511..616T,2023MNRAS.518.4672S,2025A&A...697A.211D} than cosmological simulations. A question is obviously if numerical aspects contribute to suppress MBH growth. For instance \citet{2022MNRAS.516.2112K} note that SN feedback may have been made ``super-efficient'' in order to suppress star formation in galaxies and avoid overestimating the stellar masses: as JWST is also finding more, and more massive galaxies than previous observations \citep[][and references therein]{2024arXiv240521054A}, perhaps the solution is milder SN feedback. Alternatively, since high-z galaxies appear very compact, either SN and AGN feedback is better counteracted by their deep potential wells, or stellar accretion had a more important role than expected in such extremely dense stellar environments \citep{2025ApJ...980..210Z}. For the most part cosmological simulations don't have the resolution to resolve pc-scale structures in a statistical sample of galaxies, but these are directions we should consider.    

Super-Eddington accretion remains an important topic of debate: up until last year it was frowned upon, while now it seems commonly accepted as a driver of high-z MBH growth. I've always been a great fan of super-Eddington accretion, but I feel that we still have to clarify many aspects, from the accretion disc scale (e.g., trapping, mass loss, jet production) to the galaxy scales (e.g., triggers of super-Eddington phases, effect of feedback, duration of episodes and duty-cycles). I will also note that when super-Eddington works, it naturally leads to enhanced \mbh-\mstar\ ratios \citep{2015ApJ...804..148V,2024arXiv241214248T}. 

The mystery of LRDs: are we getting closer to understanding them? The presence of dense gas, near the MBH and in a quasi-spherical configuration \citep{2025MNRAS.538.1921M,2025ApJ...980L..27I} seems to be able to explain the features of various LRDs\footnote{I recall that the definition of LRD is not univocal, and that they may represent a composite population \citep{2024ApJ...968....4P}. Uniform selection may help in identifying specific classes of sources \citep{2025arXiv250605459H}}, such as strong Balmer breaks \citep{2025arXiv250113082J,2025arXiv250316596N,2025arXiv250316600D}, X-ray weakness \citep{2025MNRAS.538.1921M} and may explain the red colors without much dust \citep{2025arXiv250707190L,2025arXiv250905434G}. Such gas envelopes \citep{2025arXiv250506965K} may have their physical origin in the quasistar model \citep{2025arXiv250709085B}. At the same time, clustering analyses \citep{2025MNRAS.539.2910P,2025ApJ...988..246M,2025arXiv250604004C} imply low galaxy and halo mass for the hosts of LRDs, which appear to be a population distinct from quasars and, if MBH masses are correct, display a very high \mbh-\mstar\ value -- even when stellar masses are inferred from environment and not SED fitting. 

Many questions are open, both theoretically and observationally. Time will tell whether the hypotheses we are making today will remain as viable options, or we have to change our picture on the formation and evolution of MBHs in the first billion years of the Universe.

\begin{acknowledgement}
I'm grateful to Alessandro Lupi for providing tabulated for data for MBH and galaxy masses that were used to create figures for this Chapter, and also for sharing his lectures on ``Astrophysical black holes: formation and evolution'', which helped shape Section~\ref{sec:dynchann}; to Monica Colpi for invaluable discussions on all topics covered in this Chapter and for conceiving the original idea for Fig.~\ref{fig:SchemeDynamics}; to Yohan Dubois for his slides on simulations; to Junyao Li for the contours in Fig.~\ref{fig:MbhMstar}; to Chris Richardson for the SEDs in Fig.~\ref{fig:SEDcompare} and to Michael Tremmel for the original idea behind Fig.~\ref{fig:SchemeModels}. I want to thank the organizers of the 54th Saas Fee school, Pascal Oesch, Romain Meyer and Michaela Hirschmann, for inviting me to give these lectures, which proved a lot of work, but also a wonderful occasion to reflect critically on old and new results, and ``connect the dots''. I am  especially thankful to Romain Meyer for reviewing this Chapter. I am also incredibly grateful to the students at the school, who showed interest and enthusiasm, and made me feel that the effort was weel worth. 

\end{acknowledgement}

\bibliographystyle{spbasic}
\bibliography{references}

\end{document}